\documentclass[preprint,12pt]{elsarticle}

\usepackage{amsmath}
\usepackage{amssymb}
\usepackage{graphicx}
\usepackage{dcolumn}
\usepackage{bm}

\begin{document}

\title{Dissipative Preparation of Many - Body Spin Steady States Using Trapped Ultracold Atoms}

\author[ust,rcnas]{Roland Cristopher F. Caballar}
\ead{rfcaballar@ust.edu.ph}
\affiliation[ust]{organization={Department of Mathematics and Physics, College of Science, University of Santo Tomas},
             addressline={Espana Boulevard, Sampaloc},
             city={Manila},
             postcode={1008},
             state={},
             country={Philippines}}
\affiliation[rcnas]{organization={Research Center for the Natural and Applied Sciences, University of Santo Tomas},
             addressline={Espana Boulevard, Sampaloc},
             city={Manila},
             postcode={1008},
             state={},
             country={Philippines}}
%




\date{\today}

\begin{abstract}
This article presents a dissipative method of creating a spin steady state, or a state whose spin expectation values approaches a fixed value over time, using a trapped gas of ultracold atoms coupled to a background BEC. The ultracold atoms are trapped in a double potential well embedded in a wide harmonic trap, which has a higher energy level than the double wells. The trapped atoms are then excited out of the double well trap into the harmonic trap using Raman lasers. Due to the coupling of the system to the background BEC, the atoms are then able to return to the double potential well by emitting an excitation into the background BEC, which serves as a reservoir of these excitations. By repeatedly coupling and uncoupling the trapped ultracold atoms and the background BEC over fixed intervals of time, the expectation value of the total spin of these atoms will, over time, reach a steady - state value.  
\end{abstract}

\maketitle

\section{Introduction}
The role of dissipation in quantum dynamics has been and continues to be an active area of research \cite{BreuerPetruccione, breuer, devega}. In particular, quantum dissipation has been used as a resource to prepare quantum states that are used in both quantum computing and quantum information \cite{verstraete,kastoryano2,kordas,reiter,nosov}. One advantage of the use of dissipative methods in quantum state preparation is that by interacting with an environment with a much larger number of degrees of freedom, a quantum system will, over time, eventually attain a steady state with regards to some physical property, thus allowing for a minimal amount of control on the part of the experimenter. 

One particular dissipative quantum state preparation system involves the use of single trapped atoms which are coupled to a reservoir and whose ground states are coupled to their excited states via Raman lasers with a given detuning and Rabi frequency. Examples of these dissipative quantums state preparation schemes are described in Refs. \cite{kastoryano,stannigel,gong,su,shao,li,kouzelis,zheng,cole,seetharam,marino}, wherein individual atoms are trapped in optical fields. The atoms are excited from their ground states to one or more of their excited states, and they decay back to their ground states via spontaneous emission of photons into the optical trap, which act as a reservoir of these photons. Through this driven - dissipative mechanism, the atom then evolves over time towards a steady state, with the steady state which it evolves to dependent on the type of atom that is trapped, as well as the trap configuration (e. g. optical lattice, optical cavity or optical tweezers). The resulting single - atom states prepared are of interest in quantum computation and quantum information. However, it is also possible to use many - body systems such as trapped bosonic or fermionic atoms or Bose - Einstein Condensates (BECs) for dissipative quantum state preparation schemes, as shown in Refs. \cite{daley,diehl,caballar,everest,iemini,caspar1,caspar2,hebenstreit,delre,sharma,ghasemian,yang}. The dissipative quantum state preparation schemes described in Refs. \cite{diehl,caballar,sharma}, in particular, are of interest because instead of using optical fields, they make use of superfluids or BECs as the bath or reservoir of excitations. This quantum state preparation scheme has the advantage of being able to prepare many - body quantum states which are of interest not just in quantum information and quantum computation, but also for more general purposes such as those produced in Refs. \cite{caspar1,caspar2,hebenstreit}, wherein the resulting steady - state of the initial many - body system consisting of spin - 1/2 particles is a BEC, and that produced in Ref. \cite{iemini}, which is a p - wave superconductor prepared using a finite number of one - dimensional fermions. Finally, it should be noted that it is also possible, as demonstrated in Ref. \cite{toth}, to formulate a dissipative quantum state preparation scheme wherein macroscopic systems such as mechanical resonators serve as reservoirs of excitations for a quantum system, in this case an ensemble of microwave photons, which results in the photons behaving coherently.

The dissipative quantum state preparation schemes mentioned above are but a sampling of many others that have been proposed and implemented over the years. One interesting application of these dissipative quantum state preparation schemes are those formulated in Refs. \cite{caspar1,caspar2,hebenstreit,iemini}, we see that it is possible to induce collective quantum behavior in the form of Bose - Einstein condensation or superconductivity via dissipative mechanisms. For BEC preparation, in particular, these results are significant, considering that the standard method of preparing BECs via optical trapping and laser cooling \cite{PethickSmith} requires that the gas of atoms be isolated from the surrounding environment to reduce the risk of thermal losses. This in turn requires a high degree of control over the BEC preparation process. However, by introducing dissipation as a dynamical resource, this significantly reduces the degree of control required of the experimenter during the process, since the dissipative dynamics can be used to drive the time evolution of the gas of atoms towards a BEC. 

Hence, motivated by these dissipative quantum state preparation mechanisms, this paper proposes a dissipative preparation mechanism using a gas of trapped ultracold atoms coupled to a background BEC which will serve as a reservoir of excitations. The mechanism proposed in this paper is based on that formulated in Ref. \cite{caballar}, which in turn is a proposed physical realization of the theoretical mechanism proposed in Ref. \cite{watanabe} for the preparation of number - and phase - squeezed states. However, instead of producing squeezed states, the dissipative preparation mechanism described in this paper will produce spin steady states, that is, states whose expectation value of their total spin remains constant over time. It is to be noted that this is not the first dissipative quantum state mechanism which will affect the spin of a many - body quantum system. Dissipative quantum state preparation systems that are capable of controlling the spin of many - body quantum systems have been proposed in Refs. \cite{seetharam,marino}. However, the significance of this many - body dissipative quantum preparation scheme is its use of a background BEC as the environment to which the many - body ultracold atom system is coupled, with interatomic interactions between the atoms comprising the trapped ultracold atom gas and the background BEC, instead of quantum electrodynamic interactions between these trapped atoms and an optical cavity, being the mechanism that enables the dissipative dynamics of the system. Such a mechanism will facilitate the preparation of quantum many - body states due to the many - body interactions between the ultracold atom gas comprising the system and the background BEC that forms the reservoir via the emission of Bogoliubov excitations from the system into the environment \cite{diehl}.

The rest of this paper proceeds as follows. In section II, we describe the components for the dissipative quantum state preparation scheme, specifying in particular the form of the system and reservoir Hamiltonians, the trapping potential to be used for the ultracold atom gas, and the interaction Hamiltonian that will describe the coupling between the trapped ultracold atom gases and the background BEC. In section III, we examine the dissipative dynamics of the system as a result of its coupling with the reservoir, deriving in particular the quantum master equation that will describe the time evolution of the trapped ultracold atom gas as it interacts with the background BEC. Section IV presents the numerical and graphical results from the implementation of this dissipative state preparation scheme, demonstrating in particular that the resulting state prepared using this mechanism will have total spin expectation values that will evolve over time to a steady state. Section V summarizes the results obtained in this paper and outlines further applications and prospects for future work. 

\section{Components of the Dissipative Quantum State Preparation Scheme}

\subsection{System and Reservoir Hamiltonians}

\begin{figure}[htb]
\includegraphics[width=1.0\columnwidth, height=0.45\textheight]{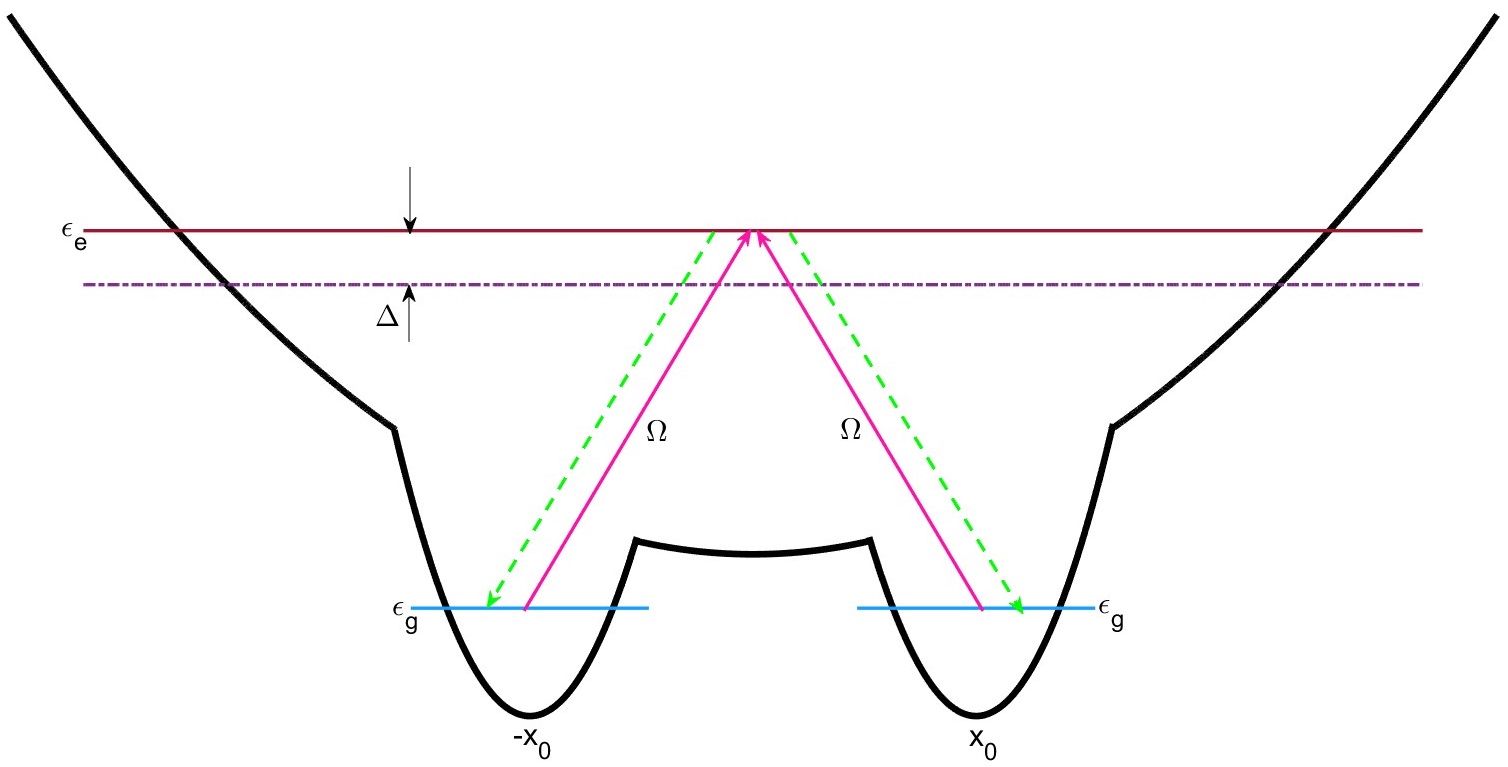}
\caption{\label{fig:sysenerglevdiag} Schematic diagram of the potential used to trap the ultracold atom gas to be used in the dissipative BEC preparation scheme. The atoms in the ground state $\epsilon_g$ of the double wells (blue lines) are coupled by Rabi lasers (pink lines) with symmetric Rabi frequencies $\Omega$ and detuning $\Delta$ (dot-dashed purple line) to the excited state $\epsilon_e$ in the wide harmonic trap (solid red line). Emission of an excitation with energy $E_\mathbf{k}$ (dashed green lines) into the background BEC will cause the excited atoms to return to the ground state in either of the two double wells.}
\end{figure}

The physical system to be used for the dissipative BEC preparation scheme is a gas of ultracold bosonic atoms, trapped in a modified double - well potential, similar to the system used in Ref. \cite{caballar}, with the schematic diagram and its corresponding energy levels shown in figure \ref{fig:sysenerglevdiag}.

As shown in the figure, the narrow double wells in the potential are centered at $x=\pm x_0$, which contain the degenerate ground states $\phi_{g,\pm}(x)$ corresponding to the energy $\epsilon_g$. The double wells, in turn, are embedded in a wide harmonic potential, which contains the excited state $\phi_{e,0}(x)$ corresponding to the energy $\epsilon_e$. The Hamiltonian for the trapped ultracold atom system can then be written in its second quantized form as
\begin{equation}
\hat{H}_{S}=\epsilon_g(\hat{a}^{\dagger}_{g,+}\hat{a}_{g,+}+\hat{a}^{\dagger}_{g,-}\hat{a}^{\dagger}_{g,-})+\epsilon_e \hat{a}^{\dagger}_{e,0}\hat{a}_{e,0},
\label{sysham}
\end{equation}
where the operators $\hat{a}_{g,\pm}$ and $\hat{a}^{\dagger}_{g,\pm}$ are the annihilation and creation operators, respectively, corresponding to the degenerate ground state energy level $\epsilon_g$ at the location of the double wells $x=\pm x_0$, and the operators $\hat{a}_{e,0}$ and $\hat{a}^{\dagger}_{e,0}$ are the annihilation and creation operators, respectively, corresponding to the non - degenerate excited energy level $\epsilon_e$ located at $x=0$. We note that the Hilbert space in which the trapped ultracold atom is described has the form $\mathcal{H}_{S}=\mathcal{H}_{S,E}\otimes\mathcal{H}_{S,p}$, where the subscripts $E$ and $p$ denote Hilbert subspaces whose basis vectors are given in terms of the eigenstates of the number and position operators $\hat{N}$ and $\hat{x}$, respectively. In particular, we can write the system operators $\hat{a}_{g,\pm}$, $\hat{a}^{\dagger}_{g,\pm}$ and $\hat{a}_{e,0}$ in the following form:
\begin{equation}
\hat{a}_{g,\pm}=\hat{a}_g \otimes\left|\pm x_0 \right>\left<\pm x_0 \right|, \hat{a}_{e,0}=\hat{a}_e \otimes\left|0\right>\left<0\right|
\label{sysannihilop}
\end{equation}
Here, $\hat{a}_g$ and $\hat{a}_e$ are the annihilation operators for the ground and excited energy states, with $\hat{a}_g \left|N_g\right>=\sqrt{N_g}\left|N_g -1\right>$ and $\hat{a}_e \left|N_e\right>=\sqrt{N_e}\left|N_e -1\right>$, where $\left|N_g\right>$ and $\left|N_e\right>$ are particle number states corresponding to the ground state and excited state of the system. On the other hand, the operators $\left|\pm x_0 \right>\left<\pm x_0 \right|$ and $\left|0 \right>\left<0 \right|$ are position operators corresponding to the locations of the double wells $x=\pm x_0$ and $x=0$ in the potential diagrammed in Fig. \ref{fig:sysenerglevdiag}.

On the other hand, the Hamiltonian for the background BEC in which the trapped ultracold atoms are immersed is given by
\begin{equation}
\hat{H}_{B}=\sum_\mathbf{k} E_\mathbf{k}\hat{b}^{\dagger}_{\mathbf{k}}\hat{b}_{\mathbf{k}}
\label{becham}
\end{equation}
In this Hamiltonian, $\hat{b}_\mathbf{k}$ and $\hat{b}^{\dagger}_\mathbf{k}$ are the annihilation and creation operators for excitations in the BEC, with those excitations having energy 
\begin{equation}
E_\mathbf{k}=\sqrt{\epsilon_k^0(\epsilon_k^0 +2\rho_b U_B)},
\label{exciener}
\end{equation}
where $\epsilon_k^0 = \frac{\hbar^2 k^2}{2m_B}$ is the energy of the excitations when they behave as free particles, and $U_B=\sqrt{\frac{4\pi\hbar^2 a_{BB}}{m_B}}$ is the interaction potential, where $m_B$ is the mass of the atoms in the background BEC, $a_{BB}$ is the scattering length of the atoms in the condensate, and $\rho_B$ is the density of atoms in the condensate. We note that the creation and annihilation operators for the system and background BEC Hamiltonians obey the following commutator relations:

\begin{eqnarray}
&&[\hat{a}_{j,n},\hat{a}_{j',n'}^{\dagger}]=\delta_{j,j'}\delta_{n,n'}, [\hat{a}_{j,n},\hat{a}_{j',n'}]=[\hat{a}^{\dagger}_{j,n},\hat{a}^{\dagger}_{j',n'}]=0\nonumber\\
&&[\hat{b}_{k},\hat{b}_{k'}^{\dagger}]=\delta_{k,k'}, [\hat{b}_k,\hat{b}_k']=[\hat{b}_{k}^{\dagger},\hat{b}_{k'}^{\dagger}]=0\nonumber\\
\label{commutatorident}
\end{eqnarray}
In the first two identities, $j=g,e$ and $n=+,0,-$. 

\subsection{Interaction Hamiltonian}

As shown in figure \ref{fig:sysenerglevdiag}, the trapped bosonic atoms interact with the BEC in which it is immersed by emitting excitations into it in order for them to return from the excited state $\epsilon_e$ in the harmonic potential to one of the degenerate ground states $\epsilon_g$ contained in the double well potential. This is done after the atoms, which are initially in the double well, are excited by Rabi lasers with identical Raman frequencies $\Omega$ and detuning $\Delta$ to the excited state in the harmonic potential. This interaction between the trapped atoms and the background BEC is described using the interaction Hamiltonian
\begin{equation}
\hat{H}_{SB}=\frac{4\pi a_{SB}}{2\mu}\int d^3 r\hat{\psi}^{\dagger}_{S}\hat{\psi}_{S}\hat{\psi}^{\dagger}_{B}\hat{\psi}_{B}
\label{interactham}
\end{equation}
In this expression, $a_{SB}$ is the inter-atomic species scattering length between the trapped ultracold atoms in the system (S) and the atoms comprising the background BEC (B), while $\mu$ is the reduced mass of the system and BEC atoms. In constructing the interaction Hamiltonian, we need the explicit form of the field operators for both the system of trapped ultracold atoms and the background BEC in which they are immersed. For the trapped ultracold atom system, the corresponding field operator has the following form:
\begin{equation}
\hat{\psi}_S = \left[\hat{a}_{g,-}\phi_{g,-}(x)+\hat{a}_{g,+}\phi_{g,+}(x)+\hat{a}_{e,0}\phi_{e,0}(x)\right]w_y (y)w_z (z)
\label{sysfieldop}
\end{equation}
In this field operator, $\phi_{g,\pm}(x)w_y(y)w_z(z)$ and $\phi_{e,0}(x)w_y(y)w_z(z)$ are the eigenfunctions corresponding to the ground state energy $\epsilon_g$ and the excited state energy $\epsilon_e$ in position representation of the trapped ultracold atom system's Hamiltonian. The field operator for the background BEC, on the other hand, can be written as
\begin{equation}
\hat{\psi}_B = \sqrt{\rho_B}+\delta\hat{\psi}_B
\label{becfieldop}
\end{equation}
where the excitation term $\delta\hat{\psi}_B$ has the explicit form
\begin{equation}
\delta\hat{\psi}_B = \frac{1}{\sqrt{V}}\sum_{\mathbf{k}}\left(u_\mathbf{k}e^{i\mathbf{k}\cdot\mathbf{r}}\hat{b}_\mathbf{k}+v_\mathbf{k}e^{-i\mathbf{k}\cdot\mathbf{r}}\hat{b}_\mathbf{k}\right),
\label{becexcifieldop}
\end{equation}
with $V$ being the total volume of the background BEC, $u_\mathbf{k}=(1-L^{2}_\mathbf{k})^{-1/2}$, $v_\mathbf{k}=L_\mathbf{k}(1-L^{2}_\mathbf{k})^{-1/2}$, $L_\mathbf{k}=(E_\mathbf{k}-(k^2/2m_{B})-m_{B}c^2)/m_{B}c^2$,  and $E_\mathbf{k}$ is the excitation energy given by Eq. \ref{exciener}.
In carrying out the integration necessary to obtain the interaction Hamiltonian for this system, we will be operating under the assumption that the excitations emitted by the ultracold atoms into the background BEC are sound - like, i. e. $E_\mathbf{k}\approx ck$, where $c=\sqrt{\frac{\rho_B U_B}{m_B}}$. Also, we will assume that the x - components of the ground state and the excited state wavefunctions for the trapped ultracold atom systems will have the following forms:

\begin{equation}
\phi_{g,\pm}(x)=\left(\frac{m_{S}\omega_{g,x}}{\pi\hbar}\right)^{1/4}\exp\left(-\frac{m_{S}\omega_{g,x}}{2\hbar}(x\pm x_0)^2\right)
\label{grndstatewavefcn}
\end{equation}
\begin{equation}
\phi_{e,0}(x)=\frac{\sqrt{2}}{\pi^{1/4}}\left(\frac{m_{S}\omega_{e,x}}{\hbar}\right)^{3/4}x\exp\left(-\frac{m_{S}\omega_{e,x}}{2\hbar}x^2\right)
\label{exctdstatewavefcn}
\end{equation}
In both equations, $\omega_{g,x}$ and $\omega_{e,x}$ are the frequencies corresponding to the ground state and excited state energies $\epsilon_g$ and $\epsilon_e$ of the system, respectively, and $m_S$ is the mass of the atoms comprising the system. Also, we assume that the transverse components $w_{y}(y)$ and $w_{z}(z)$ of the ground state and excited state wavefunctions have the same form as the x - component for the ground state wavefunction, given by Eq. \ref{grndstatewavefcn}, with $\omega_y$ and $\omega_z$ being the frequencies corresponding to these wavefunctions, and y and z replacing the x coordinate, with $y_0 = z_0 = 0$. 

Let us now proceed to the derivation of the interaction Hamiltonian itself. Substituting the explicit forms of the field operators given by equations \ref{sysfieldop} and \ref{becfieldop} into equation \ref{interactham}, we obtain
\begin{eqnarray}
&&\hat{H}_{SB}=\frac{2\pi a_{SB}}{\mu}\int d^{3}x w_{y}^{2}(y)w_{z}^{2}(z)\left\{\phi_{g,+}^{2}(x)\hat{a}_{g,+}^{\dagger}\hat{a}_{g,+}+\phi_{g,-}^{2}(x)\hat{a}_{g,-}^{\dagger}\hat{a}_{g,-}\right.\nonumber\\
&&+\phi_{e,0}^{2}(x)\hat{a}_{e,0}^{\dagger}\hat{a}_{e,0}+\phi_{g,+}(x)\phi_{e,0}(x)\left(\hat{a}_{g,+}^{\dagger}\hat{a}_{e,0}+\hat{a}_{e,0}^{\dagger}\hat{a}_{g,+}\right)\nonumber\\
&&\left.+\phi_{g,-}(x)\phi_{e,0}(x)\left(\hat{a}_{g,-}^{\dagger}\hat{a}_{e,0}+\hat{a}_{e,0}^{\dagger}\hat{a}_{g,-}\right)\right\}\nonumber\\
&&\times\left\{\rho_{B}+\sqrt{\frac{\rho_B}{V}}\sum_{k}(u_k +v_k )\left(e^{i\mathbf{k}\cdot\mathbf{r}}\hat{b}_k +e^{-i\mathbf{k}\cdot\mathbf{r}}\hat{b}^{\dagger}_k\right)\right\}\nonumber\\
\label{interactham1}
\end{eqnarray}
In carrying out the substitution, we keep only the terms that are linear in the condensate field operators $\hat{b}_k$ and $\hat{b}^\dagger _k$. For the succeeding steps, we will work in Cartesian coordinates to facilitate ease of calculation of the interaction Hamiltonian.

Using now our earlier assumption that $w_y(y)$ and $w_z(z)$ have the same form as $\phi_{g,\pm}(x)$, we can carry out the integration over the transverse variables as follows:
\begin{equation}
\int w^{2}_{u}(u)e^{\pm ik_u u}du=\sqrt{\frac{m_S \omega_{g,u}}{\pi\hbar}}\int du e^{-\frac{m_S \omega_{g,u}}{\hbar}u^2}e^{\pm ik_u u}
\end{equation}
In evaluating this integral, we can treat the ground state transverse wavefunction $w_u^2 (u)$ as a Dirac delta function by making the assumption that $\omega_{g,u}\rightarrow \infty$. Under this assumption, we then obtain the following:
\begin{equation}
\int w^{2}_{u}(u)e^{\pm ik_u u}du \rightarrow \int du \delta(u)e^{\pm ik_u u}=1
\end{equation}

Also, by making the assumption that $\omega_{g,x}\rightarrow\infty$ for the ground state wavefunctions along the x - axis, we can simplify the task of evaluating the overlap integrals $\int dx \phi_{g,\pm}(x)\phi_{e,0}(x)e^{\pm ik_x x}$ in Eq. \ref{interactham1} in the following manner:

\begin{eqnarray}
&&\int dx \phi_{g,\pm}(x)\phi_{e,0}(x)e^{\pm ik_x x}=2\sqrt{\frac{m_S \omega_{e,x}}{\hbar}}\left(\frac{\omega_{e,x}}{\omega_{g,x}}\right)^{1/4}\sqrt{\frac{m_S \omega_{g,x}}{2\pi\hbar}}\nonumber\\
&&\times\int dx e^{-\frac{m_S \omega_{g,x}}{2\hbar}(x\pm x_0)^2}e^{-\frac{m_S \omega_{e,x}}{2\hbar}x^2}e^{\pm ik_x x}x\nonumber\\
&&=\pm 2\sqrt{\frac{m_S \omega_{e,x}}{\hbar}}\left(\frac{\omega_{e,x}}{\omega_{g,x}}\right)^{1/4}e^{-\frac{m_S \omega_{e,x}}{2\hbar}x_0^2}e^{ik_x x_0}x_0\approx\pm 2\sqrt{\frac{m_S \omega_{e,x}}{\hbar}}\left(\frac{\omega_{e,x}}{\omega_{g,x}}\right)^{1/4} x_0 e^{-\frac{m_S \omega_{e,x}}{2\hbar}x_0^2} \nonumber\\
\label{overlapintapprox1}
\end{eqnarray}
In evaluating these overlap integrals, we assume that $k_x x_0<<1$ in order to make the approximations given in Eqs. \ref{overlapintapprox1}. 

It is to be noted that for the approximation $k_x x_0<<1$, mathematically, this will ensure that $\exp(ik_x x_0)\approx 1$. Physically, this means that the double wells are very narrow compared to the inter-well separation $2x_0$. This condition is necessary to ensure that there is no inter - well tunneling between the ground states. We note that the oscillator frequency for the double wells in the ground state is given by $\omega_g=\sqrt{k_x/m_s}$. The oscillator frequency, in turn, is necessary to calculate the oscillator length of the double wells, which is given by $\sigma_g = \sqrt{1/m_s \omega_g}$. Therefore, with $\omega_g\rightarrow\infty$ (a condition set earlier to ensure that the ground state wavefunctions can be approximated as Dirac delta distributions), it follows that $\sigma_g<<1$, and consequently, $\sigma_g < x_0$. However, this will imply that $k_x\rightarrow\infty$ as well, which will result in the emergence of large oscillatory terms due to $\exp(ik_x x_0)$ emerging in the sum over $k_x$. These large oscillatory terms will be problematic if we try to evaluate the sum over $k_x$, which we will need to do later when we derive the master equation. So to ensure that $k_x x_0$ is not infinite, and that no large oscillatory terms due to $\exp(ik_x x_0)$ will emerge, we can set the location of the double wells $\pm x_0$ such that their order of magnitude is $|x_0|<<1$, so that $k_x x_0$ is finite, and in particular $k_x x_0 <<1$.

At the same time, it is to be noted that for the ground state in the double - well potential and the excited state in the harmonic trap not to overlap with each other, the following condition must be satisfied:

\begin{equation}
\omega_{g,x} < 3\omega_{e,x}
\label{freqcond}
\end{equation}

As such, any shift in the ground state energy in the double - well potential must also result in a corresponding shift in the energy of the excited state in the harmonic trap to which the ground states in the double - well potential are coupled, with the magnitude of that shift being $\Delta\omega_{e,x}\geq 2\omega_{g,x}$. 

Finally, using the same assumptions as those given in the previous paragraphs regarding $\omega_{g,x}$ and $\omega_{e,x}$, the overlap integrals $\int dx \left|\phi_{g,\pm}(x)\right|^2 e^{\pm ik_x x}$ and $\int dx \left|\phi_{e,0}(x)\right|^2 e^{\pm ik_x x}$ can be evaluated as follows:

\begin{eqnarray}
&&\int dx \left|\phi_{g,\pm}(x)\right|^2 e^{\pm ik_x x}\rightarrow \int dx \delta(x\pm x_0)e^{\pm ik_x x}=e^{\pm ik_x x_0}\approx 1\nonumber\\
&&\int dx \left|\phi_{e,0}(x)\right|^2 e^{\pm ik_x x}\rightarrow 2\left(\frac{m_S \omega_{e,x}}{\hbar}\right) \int dx \delta(x)x^2 e^{\pm ik_x x}=0\nonumber\\
\end{eqnarray}
 
Combining all of these results, we obtain the following form of the interaction Hamiltonian:

\begin{eqnarray}
&&\hat{H}_{SB}=\frac{2\pi a_{SB}}{\mu}\left(\rho_{B}\left(\hat{a}_{g,+}^{\dagger}\hat{a}_{g,+}+\hat{a}_{g,-}^{\dagger}\hat{a}_{g,-}+\hat{a}_{e,0}^{\dagger}\hat{a}_{e,0}\right)\right.\nonumber\\
&&\left.+\sqrt{\frac{\rho_B}{V}}\sum_{k}(u_k +v_k )\left(\hat{a}_{g,+}^{\dagger}\hat{a}_{g,+}+\hat{a}_{g,-}^{\dagger}\hat{a}_{g,-}\right)(\hat{b}_k +\hat{b}_k^\dagger )\right)\nonumber\\
&&+\frac{4\pi a_{SB}}{\mu}\sqrt{\frac{m_S \omega_{e,x}\rho_B}{V\hbar}}\left(\frac{\omega_{e,x}}{\omega_{g,x}}\right)^{1/4}x_0 e^{-\frac{m_s\omega_e}{2\hbar} x_0^2}\nonumber\\
&&\times\sum_{k}(u_k +v_k )\left((\hat{a}^{\dagger}_{g,+}-\hat{a}^{\dagger}_{g,-})\hat{a}_{e,0}+(\hat{a}_{g,+}-\hat{a}_{g,-})\hat{a}^{\dagger}_{e,0}\right)(\hat{b}_k +\hat{b}_k^\dagger)\nonumber\\
\label{interacthamexp1}
\end{eqnarray}

\subsection{Simplification of the Interaction Hamiltonian}
Now let us simplify this interaction Hamiltonian by first making use of the explicit form of the coefficients $u_k$ and $v_k$.  We will be working in the phonon limit, i. e. in the limit where the excitations have energy $E_k = ck$, where, again, $c=\sqrt{\frac{\rho_B U_B}{m_B}}$. In the phonon limit, $k$ is very small, and in particular, $k<<m_B c$ so that 
\begin{equation}
L_\mathbf{k}=(E_\mathbf{k}-(k^2/2m_{B})-m_{B}c^2)/m_{B}c^2\approx \frac{k}{m_B c}-1
\end{equation}
As such, evaluating the sum of $u_k$ and $v_k$, we obtain
\begin{eqnarray}
&&u_k +v_k = \frac{1+L_k}{\sqrt{1-L^{2}_k}}=\sqrt{\frac{1+L_k}{1-L_k}}=\sqrt{\frac{\frac{k}{2m_B c}}{2-\frac{k}{2m_B c}}}\nonumber\\
&&\frac{1}{2}\sqrt{\frac{k}{m_B c}}\left(1-\frac{k}{4m_B c}\right)^{-1/2}\approx\frac{1}{2}\sqrt{\frac{k}{m_B c}}\nonumber\\
\label{coeffsum1}
\end{eqnarray}
Substituting this in Eq. \ref{interacthamexp1}, we then obtain
\begin{eqnarray}
&&\hat{H}_{SB}=\frac{2\pi a_{SB}}{\mu}\left(\rho_{B}(\hat{a}_{g,+}^{\dagger}\hat{a}_{g,+}+\hat{a}_{g,-}^{\dagger}\hat{a}_{g,-}+\hat{a}_{e,0}^{\dagger}\hat{a}_{e,0})\right.\nonumber\\
&&+\frac{1}{2}\sqrt{\frac{\rho_B}{Vm_B c}}\sum_{k}\sqrt{k}(\hat{a}_{g,+}^{\dagger}\hat{a}_{g,+}+\hat{a}_{g,-}^{\dagger}\hat{a}_{g,-})(\hat{b}_k +\hat{b}_k^\dagger )\nonumber\\
&&+\sqrt{\frac{m_S \omega_{e,x}\rho_B}{V\hbar m_B c}}\left(\frac{\omega_{e,x}}{\omega_{g,x}}\right)^{1/4}x_0 e^{-\frac{m_S \omega_{e,x}}{2\hbar}x_0^2}\sum_{k}\sqrt{k}\left((\hat{a}^{\dagger}_{g,+}-\hat{a}^{\dagger}_{g,-})\hat{a}_{e,0}+(\hat{a}_{g,+}-\hat{a}_{g,-})\hat{a}^{\dagger}_{e,0}(\hat{b}_k +\hat{b}_k^\dagger))\right)\nonumber\\
\label{interacthamexp2}
\end{eqnarray}
\section{Dynamics of the Driven - Dissipative BEC Preparation Scheme}

\subsection{Time Evolved Interaction Hamiltonian}

The interaction Hamiltonian is evolved over time by applying the Baker - Campbell - Hausdorff (BCH) identity, whose explicit form is given as \cite{sakurai}
\begin{equation} 
\hat{A}(t)=\exp\left(\frac{it}{\hbar}\hat{H}\right)\hat{A}\exp\left(-\frac{it}{\hbar}\hat{H}\right)=\hat{A}+\frac{it}{\hbar}\left[\hat{H},\hat{A}\right]+\frac{1}{2!}\left(\frac{it}{\hbar}\right)^{2}\left[\hat{H},\left[\hat{H},\hat{A}\right]\right]+...
\label{BCHident}
\end{equation}
In doing so, the time - evolved system and background BEC annihilation operators will then have the following form:
\begin{equation}
\hat{a}_{j,n}(t)=e^{\frac{it}{\hbar}\hat{H}_S}\hat{a}_{j,n}e^{-\frac{it}{\hbar}\hat{H}_S}=e^{-\frac{it}{\hbar}\epsilon_g}\hat{a}_{j,n},\hat{b}_{k}(t)=e^{\frac{it}{\hbar}\hat{H}_B}\hat{b}_{k}e^{-\frac{it}{\hbar}\hat{H}_B}=e^{-\frac{it}{\hbar}E_k}\hat{b}_{k}
\label{timeevosysop}
\end{equation}
The time - evolved system and background BEC creation operators, on the other hand, are obtained by taking the Hermitian conjugate of the time - evolved system and background BEC annihilation operators. 

Substituting these time - evolved creation and annihilation operators in the interaction Hamiltonian, we then obtain the following explicit form of the time - evolved interaction Hamiltonian:
\begin{eqnarray}
&&\hat{H}_{SB}(t)=\frac{\pi a_{SB}}{\mu}\sqrt{\frac{\rho_B}{Vm_B c}}\sum_{k}\sqrt{k}\left((\hat{a}_{g,+}^{\dagger}\hat{a}_{g,+}+\hat{a}_{g,-}^{\dagger}\hat{a}_{g,-})\right.\nonumber\\
&&\times(e^{-\frac{it}{\hbar}E_k}\hat{b}_k +e^{\frac{it}{\hbar}E_k}\hat{b}_k^\dagger)\nonumber\\
&&+2\sqrt{\frac{m_S \omega_{e,x}}{\hbar}}\left(\frac{\omega_{e,x}}{\omega_{g,x}}\right)^{1/4}x_0 e^{-\frac{m_S \omega_{e,x}}{2\hbar}x_0^2}\nonumber\\
&&\times\left(e^{-\frac{it}{\hbar}(E_k +(\epsilon_e -\epsilon_g))}(\hat{a}^{\dagger}_{g,+}-\hat{a}^{\dagger}_{g,-})\hat{a}_{e,0}\hat{b}_k+e^{\frac{it}{\hbar}(E_k -(\epsilon_e -\epsilon_g))}(\hat{a}^{\dagger}_{g,+}-\hat{a}^{\dagger}_{g,-})\hat{a}_{e,0}\hat{b}^{\dagger}_k\right.\nonumber\\
&&\left.+e^{-\frac{it}{\hbar}(E_k -(\epsilon_e -\epsilon_g))}(\hat{a}_{g,+}-\hat{a}_{g,-})\hat{a}^{\dagger}_{e,0}\hat{b}_k+e^{\frac{it}{\hbar}(E_k +(\epsilon_e -\epsilon_g))}(\hat{a}_{g,+}-\hat{a}_{g,-})\hat{a}^{\dagger}_{e,0}\hat{b}^{\dagger}_k)\right)\nonumber\\
\label{interacthamevo}
\end{eqnarray}

We note that in deriving Eq. \ref{interacthamevo} by time - evolving the interaction Hamiltonian given by Eq. \ref{interacthamexp2} via the BCH identity, the first term in Eq. \ref{interacthamexp2} vanishes. This is because this term commutes with the first two terms of the system Hamiltonian as well as the environment Hamiltonian, given by Eqs. \ref{sysham} and \ref{becham} respectively. Explicitly, 

\begin{eqnarray}
&&\left[\hat{H}_S\otimes\hat{H}_B,\rho_{B}(\hat{a}_{g,+}^{\dagger}\hat{a}_{g,+}+\hat{a}_{g,-}^{\dagger}\hat{a}_{g,-}+\hat{a}_{e,0}^{\dagger}\hat{a}_{e,0})\right]\nonumber\\
&&=\left[\hat{H}_S,\hat{a}_{g,+}^{\dagger}\hat{a}_{g,+}+\hat{a}_{g,-}^{\dagger}\hat{a}_{g,-}+\hat{a}_{e,0}^{\dagger}\hat{a}_{e,0}\right]\otimes\left[\hat{H}_B, \rho_B\right]\nonumber\\
&&=\left[\epsilon_g(\hat{a}^{\dagger}_{g,+}\hat{a}_{g,+}+\hat{a}^{\dagger}_{g,-}\hat{a}^{\dagger}_{g,-})+\epsilon_e \hat{a}^{\dagger}_{e,0}\hat{a}_{e,0},\hat{a}_{g,+}^{\dagger}\hat{a}_{g,+}+\hat{a}_{g,-}^{\dagger}\hat{a}_{g,-}+\hat{a}_{e,0}^{\dagger}\hat{a}_{e,0}\right]\nonumber\\
&&\otimes\sum_\mathbf{k} E_\mathbf{k}\left[\hat{b}^{\dagger}_{\mathbf{k}}\hat{b}_{\mathbf{k}},\rho_B\right]\nonumber\\
&&=\epsilon_g\left[(\hat{a}^{\dagger}_{g,+}\hat{a}_{g,+}+\hat{a}^{\dagger}_{g,-}\hat{a}^{\dagger}_{g,-}),\hat{a}_{g,+}^{\dagger}\hat{a}_{g,+}+\hat{a}_{g,-}^{\dagger}\hat{a}_{g,-}\right]+\epsilon_e\left[\hat{a}^{\dagger}_{e,0}\hat{a}_{e,0},\hat{a}_{e,0}^{\dagger}\hat{a}_{e,0}\right]\nonumber\\
&&\otimes\sum_\mathbf{k} E_\mathbf{k}\left(N_{\mathbf{k}}\rho_B-\rho_B N_{\mathbf{k}}\right)=0\nonumber\\
\end{eqnarray} 

Here, $N_\mathbf{k}$ is the eigenvalue of the number operator $\hat{N}_\mathbf{k}=\hat{b}_{\mathbf{k}}^{\dagger}\hat{b}_\mathbf{k}$, and where we used the number state representation of the environment density matrix, $\rho_B = \sum_{\mathbf{k}}c_\mathbf{k}\left|N_\mathbf{k}\right\rangle\left\langle N_\mathbf{k}\right|$.

\subsection{Derivation of the Master Equation}

Having obtained the time - evolved interaction Hamiltonian, we now proceed to derive the master equation which governs the time evolution of the system. To do so, we make use of the Born - Markov approximation, under which the master equation has the form
\begin{equation}
\frac{d\hat{\rho}}{dt}=\mathcal{L}\hat{\rho}=-\int_{0}^{\infty}dt'\:\text{Tr}_{\hat{R}}[\hat{H}_{SB}(t),[\hat{H}_{SB}(t-t'),\hat{\rho}\otimes\hat{R}]]
\label{masteq}
\end{equation}
Details surrounding the derivation of this equation can be found in Ref. \cite{BreuerPetruccione}. In this equation, $\hat{R}$ is the density matrix corresponding to the background BEC, $\hat{\rho}$ is the density matrix corresponding to the trapped ultracold atom system, and $\hat{H}_{SB}(t)$ is the time - evolved Hamiltonian describing the interaction between the system and the environment. We note that in deriving this master equation, we assume that the coupling between the system and the environment is weak, and that the time scale over which the system varies is much larger compared to the time scale over which the environment correlation functions decay. Let us now substitute Eq. \ref{interacthamevo} into this equation, and evaluate the commutators and the integral accordingly. In doing so, we obtain terms proportional to $\text{Tr}(\hat{b}_k \hat{b}_k' \hat{R})$ and $\text{Tr}(\hat{b}_k^\dagger \hat{b}_k'^\dagger \hat{R})$, both of which are equal to zero, following Refs. \cite{caballar} and \cite{daley}. At the same time, we will also be obtaining terms proportional to $\text{Tr}(\hat{b}_k^\dagger \hat{b}_k' \hat{R})$ and $\text{Tr}(\hat{b}_k \hat{b}_k'^\dagger \hat{R})$. Following Ref. \cite{caballar}, under the assumption that the background BEC has a temperature $T\approx 0$, $\text{Tr}(\hat{b}_k^\dagger \hat{b}_k' \hat{R})\approx 0$ and $\text{Tr}(\hat{b}_k \hat{b}_k'^\dagger \hat{R})\approx \delta_{k,k'}$, so that we will only consider terms in the master equation that are proportional to the latter expression. The explicit forms of these expressions are given by Eqs. \ref{masteqterm1prov} to \ref{masteqterm4prov} in the Appendices of this paper.

Next, as per Eq. \ref{masteq}, we integrate Eqs. \ref{masteqterm1prov} to \ref{masteqterm4prov} over time and sum them over $k$. In doing so, we will have to evaluate integrals of the form $\sum_{k}k\int_{0}^{\infty}dt'\;e^{\pm\frac{it'}{\hbar}E_k}$ and $\sum_{k}\int_{0}^{\infty}dt\;e^{\pm\frac{it'}{\hbar}(E_{k}\pm(\epsilon_e -\epsilon_g))}$. To evaluate these terms, let us first recall that the excitations emitted by the trapped ultracold atom gas into the background BEC are phonons with energies $E_k = ck$. At the same time, we can replace the summation over $k$ with an integration over the same variable, treating it as continuous instead of discrete. Finally, considering that the integrals over $t'$ are oscillatory integrals, we can treat them as Dirac delta distributions over $k$. Taking these all together, we obtain the following expressions:
\begin{equation}
\sum_{k}k\int_{0}^{\infty}dt'\;e^{\pm\frac{it'}{\hbar}E_k}=\int dk\;k\;\delta\left(\frac{c}{\hbar}k\right)=0
\label{oscinteg1b}
\end{equation}
\begin{equation}
\sum_{k}k\int_{0}^{\infty}dt\;e^{\pm\frac{it'}{\hbar}(E_{k}\pm(\epsilon_e -\epsilon_g))}=\int dk\;k\;\delta\left(\frac{c}{\hbar}k\pm(\epsilon_e -\epsilon_g)\right)=\mp\frac{\hbar}{c}(\epsilon_e -\epsilon_g)
\label{oscinteg2b}
\end{equation}
Substituting these terms into Eqs. \ref{masteqterm1prov} to \ref{masteqterm4prov} will result in the elimination of terms in the master equation which are proportional to the oscillatory integral $\int_{0}^{\infty}dt'\;e^{\pm\frac{it'}{\hbar}ck}$. This leaves us with terms which are proportional to the oscillatory integral $\int_{0}^{\infty}dt\;e^{\pm\frac{it'}{\hbar}(ck\pm(\epsilon_e -\epsilon_g))}$. Further simplification of the master equation can be made by making the assumption that $4\frac{m_S \omega_{e,x}}{\hbar} x^{2}_0 >>2\sqrt{\frac{m_S \omega_{e,x}}{\hbar}}x_0$. 

Finally, we perform an adiabatic elimination of the excited states in order to express them as a linear combination of the ground states trapped in the two wells in Fig. \ref{fig:sysenerglevdiag}, which can be done by ensuring that the Raman lasers coupling the ground states in the double wells to the excited state in the harmonic trap are both weak and far detuned. In so doing, and by noting that both Raman lasers have the same frequency $\Omega$, we obtain the following expression for the excited state annihilation operator $\hat{a}_{e,0}$:
\begin{equation}
\hat{a}_{e,0}\approx\frac{\Omega}{\sqrt{2}\Delta}\left(\hat{a}_{g,+}+\hat{a}_{g,-}\right)
\label{excopadiabel}
\end{equation}
We then substitute this expression for $\hat{a}_{e,0}$, together with its Hermitian conjugate, into the master equation, and group together like terms in the equation. In doing so, the master equation will have the following form:
\begin{equation}
\frac{d\hat{\rho}}{dt}\approx \gamma\left(2(\hat{c}\hat{\rho}\hat{c}^{\dagger}-\left\{\hat{\rho},\hat{c}^{\dagger}\hat{c}\right\})-(2\hat{c}^\dagger\hat{\rho}\hat{c}-\left\{\hat{\rho},\hat{c}\hat{c}^{\dagger}\right\})+e^{-\frac{2it}{\hbar}(\epsilon_e -\epsilon_g)}\left[\hat{\rho},\hat{c}\hat{c}\right]-e^{\frac{2it}{\hbar}(\epsilon_e -\epsilon_g)}\left[\hat{\rho},\hat{c}^{\dagger}\hat{c}^{\dagger}\right]\right)
\label{masteqevo4}
\end{equation}
In this equation, the jump operator $\hat{c}$ has the explicit form
\begin{equation}
\hat{c}=(\hat{a}^{\dagger}_{g,+}-\hat{a}^{\dagger}_{g,-})(\hat{a}_{g,+}+\hat{a}_{g,-})
\label{jumpop}
\end{equation}
and the coefficient $\gamma$, also known as the coupling coefficient since it describes the strength of coupling or interaction between the system and the environment, has the form
\begin{equation}
\gamma=4N_B \Omega\frac{m_S \omega_{e,x}}{\Delta}\sqrt{\frac{\omega_{e,x}}{2\omega_{g,x}}}\left(\frac{\pi a_{SB}}{\mu Vc}x_0\right)^2 e^{-\frac{m_S \omega_{e,x}}{\hbar}x_0^2}(\epsilon_e -\epsilon_g)
\label{gammacoeff}
\end{equation}

\section{Numerical Results}

\subsection{Time Evolution of the System}

Let us now carry out the time evolution of the system using the master equation given by Eq. \ref{masteqevo4}. For our initial state, it is given by the density matrix $\hat{\rho}_{0}=\left|\psi_{0}\right>\left<\psi_{0}\right|$, where the initial state ket $\left|\psi_0 \right>$ will have the form
\begin{equation}
\left|\psi_{0}\right>=\frac{1}{\sqrt{u}}\sum_{j=1}^{u}\left|N_{j,g}\right>\otimes\left|x_0\right>+\frac{1}{\sqrt{v}}\sum_{j=1}^{v}\left|N_{j,g}\right>\otimes\left|-x_0\right>
\label{initstate}
\end{equation}
Here, $\left|N_{j,g,\pm}\right>=\left|N_{j,g}\right>\otimes\left|\pm x_0\right>$ are eigenstates of the operator $\hat{N}_{g,\pm}=\hat{N}_{g}\otimes\left|\pm x_0\right\rangle\left\langle\pm x_0\right|$, where $\hat{N}_g$ is the particle number operator for the ground state. $\hat{N}_{g,\pm}$ represents a measurement of the number of particles in one of the double wells located at $x=\pm x_0$. These eigenstates have corresponding eigenvalues $N_{j,g,pm}$, and $u,v\leq N$ is the number of particle number states corresponding to the double well located at $x=\pm x_0$, respectively in the initial state. We then substitute the initial state given by Eq. \ref{initstate} into the master equation given by Eq. \ref{masteqevo4}, in order to evolve the state over time. 

\subsection{Spin Steady State Formation}

\begin{figure}[htb]
\includegraphics[width=0.5\columnwidth, height=0.2\textheight]{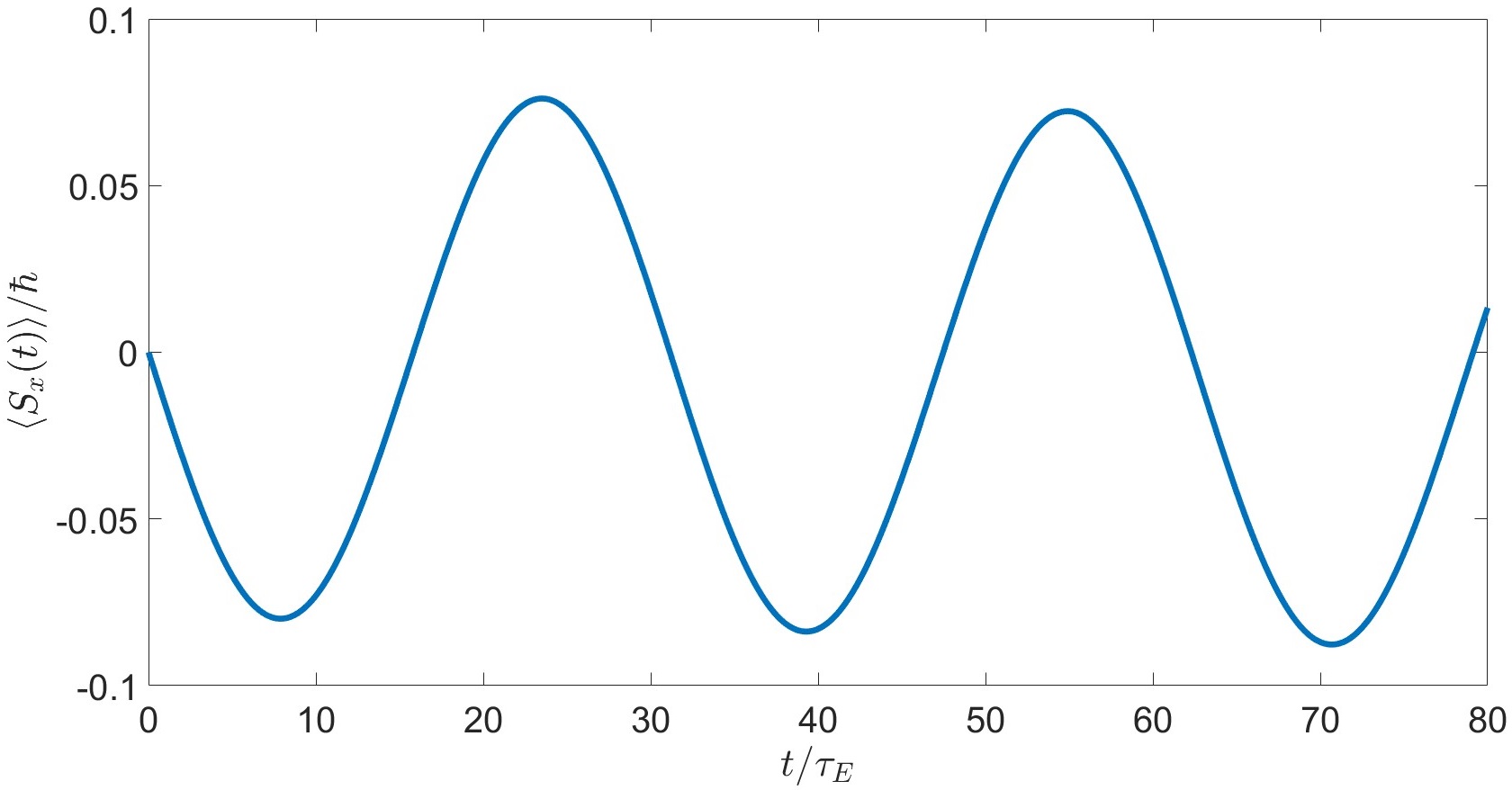}
\includegraphics[width=0.5\columnwidth, height=0.2\textheight]{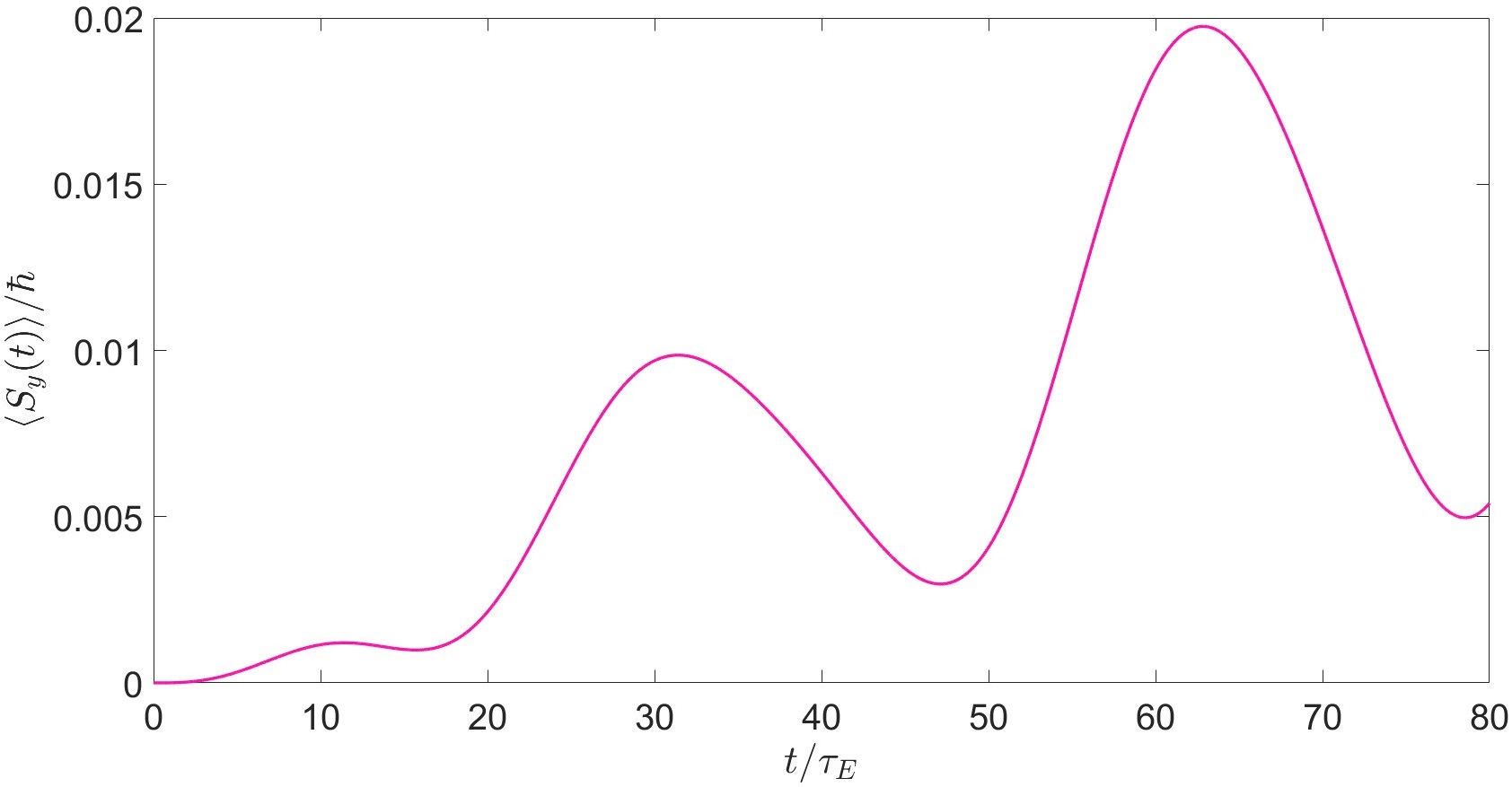}
\includegraphics[width=0.5\columnwidth, height=0.2\textheight]{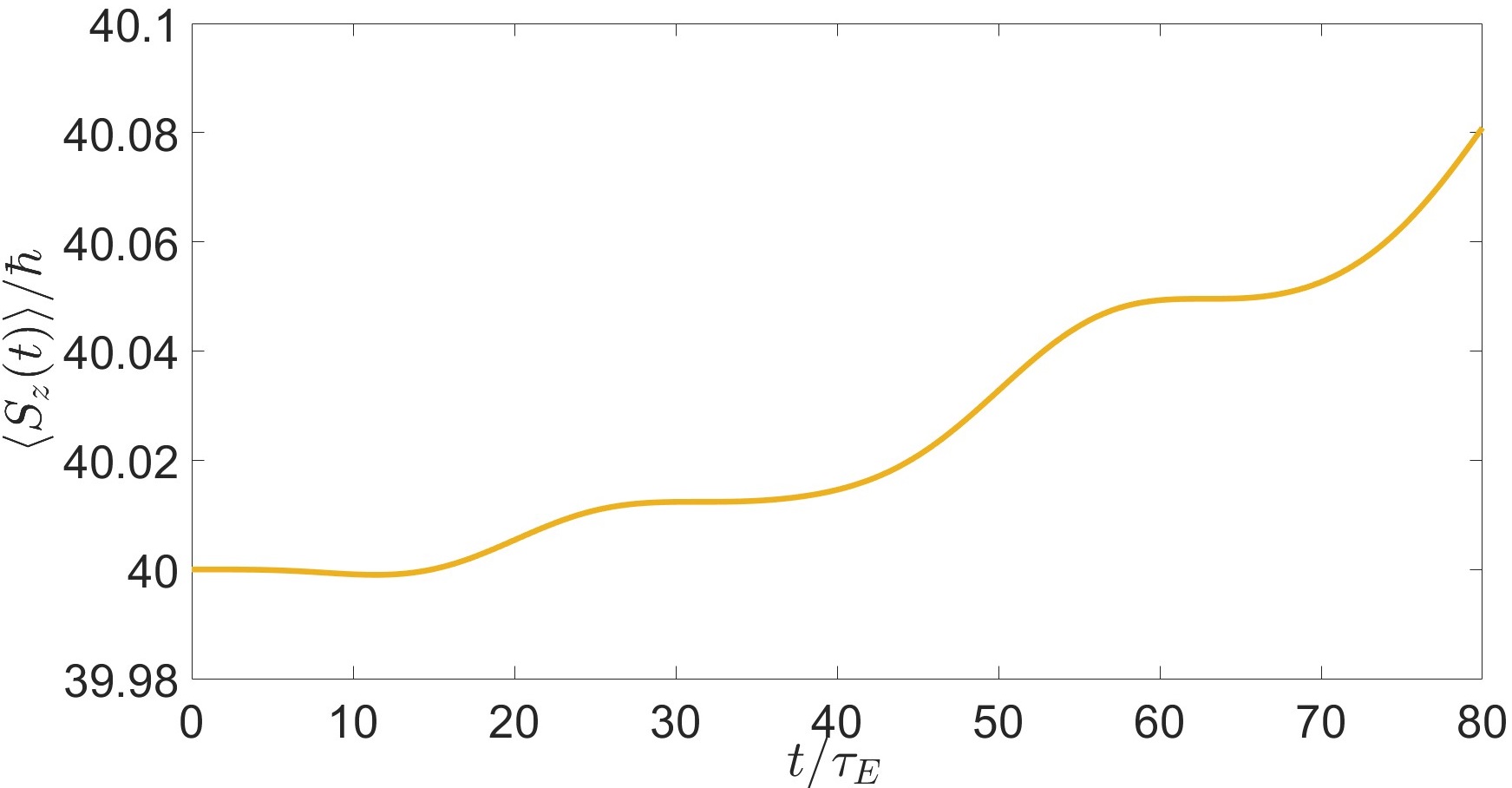}
\caption{\label{fig:spinexpvaldiag} Time evolution of the expectation values of the $SU(2)$ generators $\hat{S}_x (t)$ (top), $\hat{S}_y (t)$ (middle) and $\hat{S}_z$ (bottom), calculated using the time - evolved states $\left|\psi(t)\right\rangle$, with $u=v=2$ for these states and $\gamma$ remaining constant over the time evolution of the system.}
\end{figure}

As the trapped ultracold atom gases evolve over time, we calculate the expectation values of the $SU(2)$ generators which describe the spin-x, spin-y and spin-z of these states. These $SU(2)$ generators have the following form, as specified in Ref. \cite{watanabe} and \cite {caballar}, have the form
\begin{eqnarray}
&&\hat{S}_x = \frac{\hbar}{2}\left(\hat{a}^{\dagger}_{g,+}\hat{a}_{g,-}+\hat{a}^{\dagger}_{g,-}\hat{a}_{g,+}\right), \hat{S}_y = -\frac{i\hbar}{2}\left(\hat{a}^{\dagger}_{g,+}\hat{a}_{g,-}-\hat{a}^{\dagger}_{g,-}\hat{a}_{g,+}\right),\nonumber\\ 
&&\hat{S}_z = \frac{\hbar}{2}\left(\hat{a}^{\dagger}_{g,+}\hat{a}_{g,+}-\hat{a}^{\dagger}_{g,-}\hat{a}_{g,-}\right)\nonumber\\
\label{spinop} 
\end{eqnarray}
The resulting expectation values of these $SU(2)$ generators will have the following explicit form:
\begin{equation}
\left\langle \hat{S}_{j}(t)\right\rangle = \left\langle\psi(t)\right|\hat{S}_{j}\left|\psi(t)\right\rangle, j=x,y,z.
\label{expvalspin}
\end{equation}
We consider a system where $N=200$, $\Delta\epsilon=\epsilon_e - \epsilon_g = 1.0\times 10^{-1} J$ and $\gamma=1.0\times 10^{-5}$, where we set $\hbar=1$. The time scale over which the system is evolved is $\tau_E=\hbar / \Delta \epsilon$. The resulting expectation values for the $SU(2)$ generators $\hat{S}_j$ are shown in figure \ref{fig:spinexpvaldiag}. For this and subsequent figures, these expectation values are plotted in units of $\left\langle\hat{S}_j/\right\rangle\hbar$, $j=x,y,z$. 

As can be seen in the figure, both the expectation values of $\hat{S}_y$ and $\hat{S}_z$ exhibit very small variations over time compared to the variations shown by the expectation value of $\hat{S}_x$. Furthermore, the expectation value of $\hat{S}_x$ oscillates over time, with an amplitude much larger than the expectation values of $\hat{S}_y$ and $\hat{S}_z$. As such, for the trapped ultracold atom gases to be steady states with respect to $\hat{S}_j$, there is a need to evolve the system in such a way that $\hat{S}_x$ will approach a fixed value as the system evolves over time. 

\begin{figure}[htb]
\includegraphics[width=1.0\columnwidth, height=0.3\textheight]{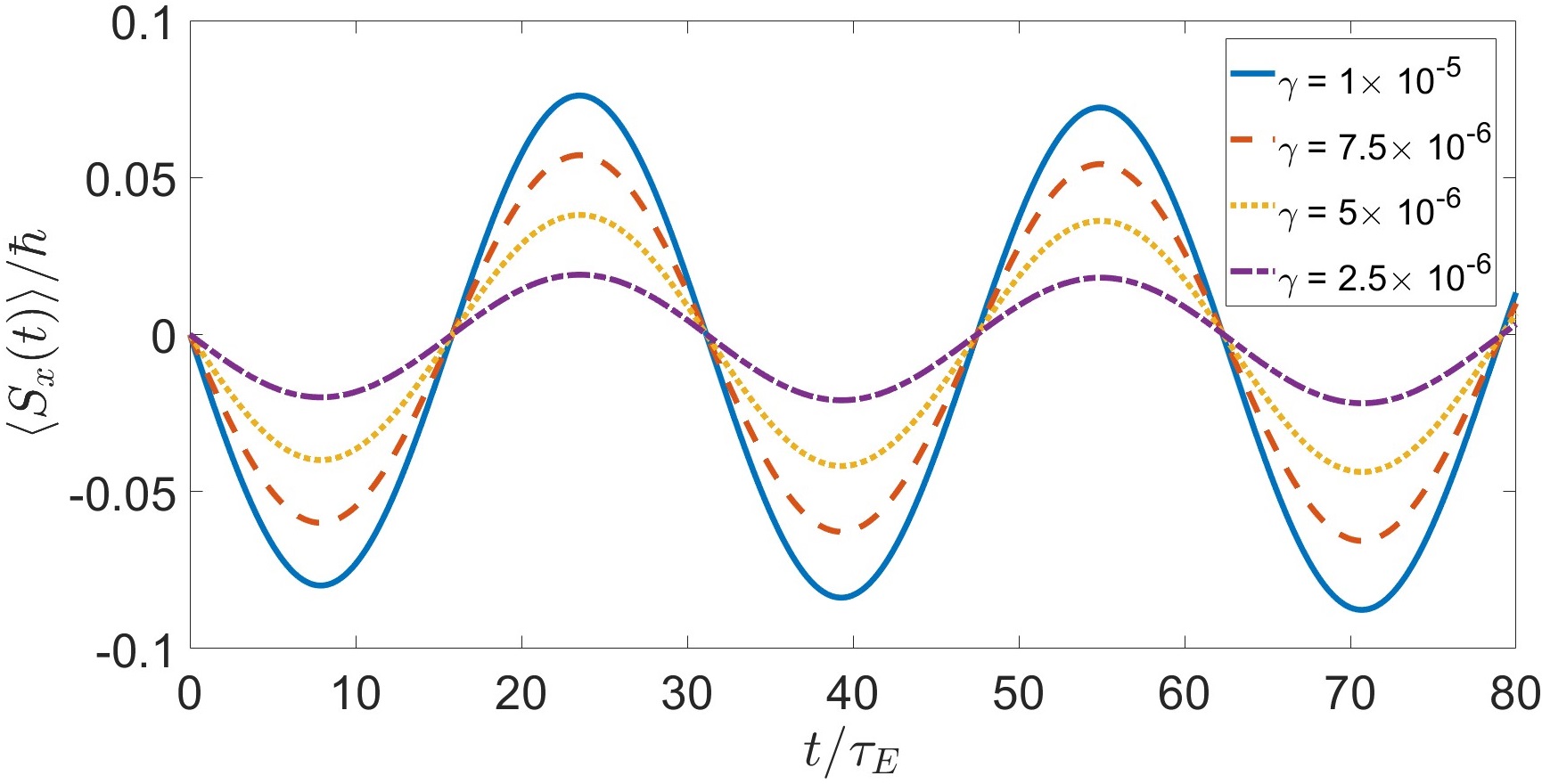}
\caption{\label{fig:spinxexpvaldecreasegamma} Time - evolved expectation value of the $SU(2)$ generator $\hat{S}_x (t)$, for decreasing values of $\gamma$, which remains constant over the time evolution of the system.}
\end{figure}

One way to achieve this is by noting that for a fixed value of $\Delta\epsilon = \epsilon_e - \epsilon_g$, the amplitude of oscillations for $\hat{S}_x$ decrease as the coupling coefficient $\gamma$, whose explicit form is given by Eq. \ref{gammacoeff}, decreases in magnitude, as shown in fig. \ref{fig:spinxexpvaldecreasegamma}. However, it should be noted that decreasing the value of $\gamma$ alone while keeping it constant throughout the time evolution of the system will not cause the system to evolve towards a steady - state value of $\hat{S}_x$, since the oscillatory behavior will still remain, albeit with a reduced amplitude. Instead, what can be done is to evolve the trapped ultracold atom gas in a stroboscopic manner, similar to the dissipative quantum state preparation scheme described in Refs. \cite{gong,caballar}. To do this, we first evolve the trapped ultracold atom gas over a time interval $\Delta t_1=t_1$ for an initial value of the coupling constant $\gamma=\gamma_1$, then turn off the coupling between the system and the environment (i. e. set $\gamma=0$) for an interval of time $\tau_1<<t_1$. We then turn the coupling between the system and the environment back on again as the system evolves over a time interval $\Delta t_2=t_2-(t_1+\tau_1)$. However, unlike what was proposed in Refs. \cite{gong,caballar}, instead of the coupling constant remaining constant, for this time interval, the coupling constant is now reduced from $\gamma=\gamma_1$ to $\gamma_2=\gamma_1 \Delta\gamma$, where $\Delta\gamma=\gamma_{n+1}/\gamma_n <1$ for $n\geq 1$. We then repeat the process multiple times until the amplitude of the oscillations have been reduced significantly such that the expectation value of $\hat{S}_x$ for the time - evolved state is almost constant. The result is a state whose expectation value of $\hat{S}_x$ evolves towards a steady state, with the oscillations of this expectation value being dampened as a result of the successive decoupling and coupling (with ever - decreasing strength) of the system with the environment. 

\begin{figure}[htb]
\includegraphics[width=1.0\columnwidth, height=0.3\textheight]{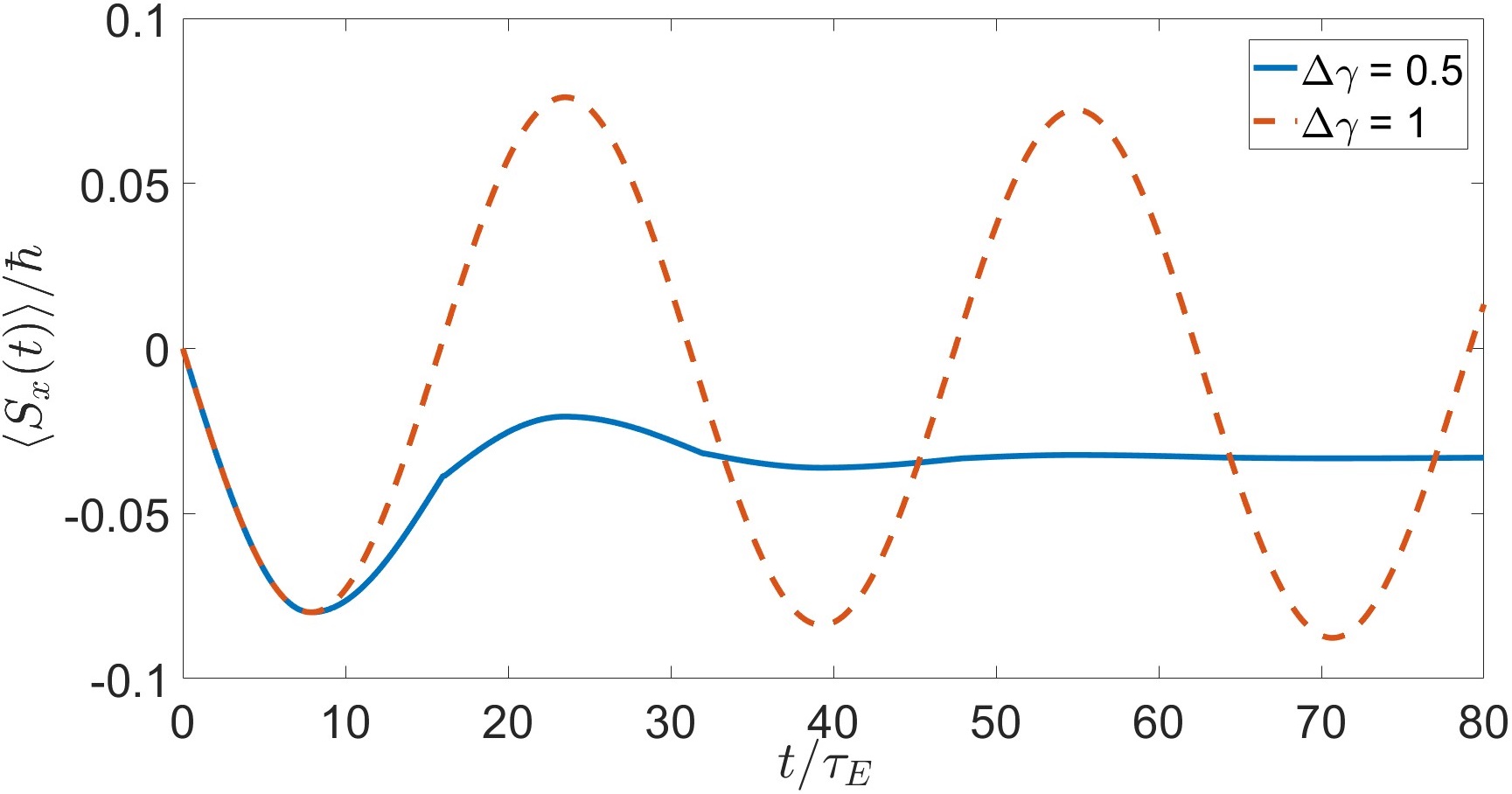}
\caption{\label{fig:spinxexpvalvaryingvsconstgamma} Time - evolved expectation value of the $SU(2)$ generator $\hat{S}_x (t)$, with $\gamma$ varying over time as $\gamma_{n}=0.5\gamma_{n-1}$, $\gamma_1=1\times 10^{-5}$, $\tau_1 = 0.1\;s$, $t_1 = 8.0\;s$, $\Delta t_n = t_n - (t_{n-1}+\tau_1)$, $n\geq 1$ (solid line) and $\gamma=1\times 10^{-5}$ remaining constant over the time evolution of the system (dashed line).}
\end{figure}

To illustrate this, we compare the time evolution of the system with and without stroboscopic coupling and decoupling with the environment. As shown in figure \ref{fig:spinxexpvalvaryingvsconstgamma}, we can see that using the stroboscopic method of time evolution outlined above, the oscillations of $\left\langle\hat{S}_x (t)\right\rangle$ are continuously dampened over each interval of time $\Delta t_{n}$ during which the system is evolved while coupled to the BEC excitation reservoir, with $\left\langle\hat{S}_x (t)\right\rangle$ approaching a steady - state value as $t$ increases. Contrast this with the case wherein the coupling constant of the system with the environment remains constant. If the system is evolved over time in this manner, then $\left\langle\hat{S}_x (t)\right\rangle$ will not approach a steady - state value due to its oscillatory behavior.

\begin{figure}[htb]
\includegraphics[width=1.0\columnwidth, height=0.3\textheight]{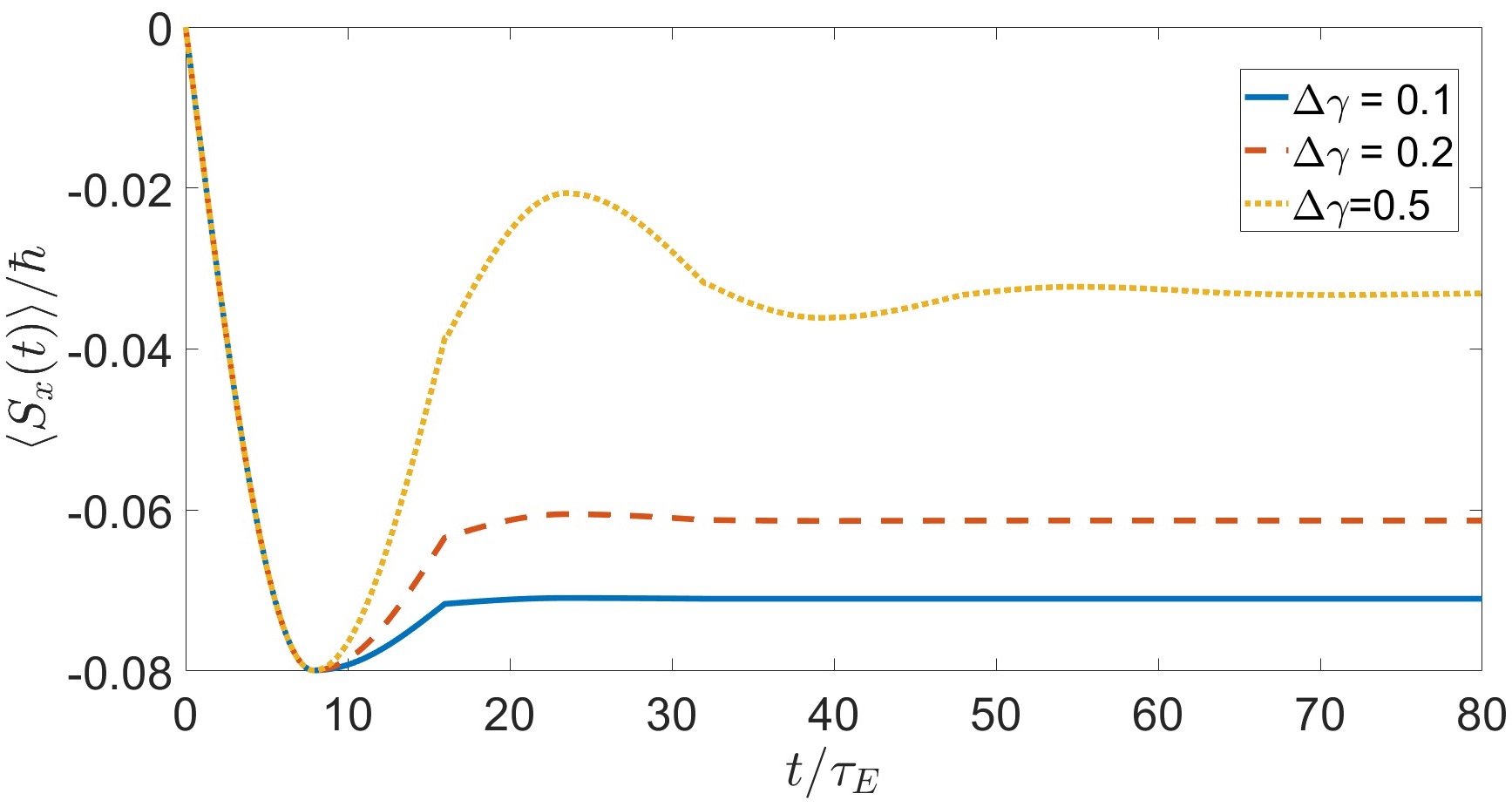}
\caption{\label{fig:spinxexpvalvaryinggamma} Time - evolved expectation value of the $SU(2)$ generator $\hat{S}_x (t)$, with $\gamma$ varying over time as $\gamma_{n}=0.1\gamma_{n-1}$ (solid line), $\gamma_{n}=0.2\gamma_{n-1}$ (dashed line), and $\gamma_{n}=0.5\gamma_{n-1}$ (dotted line), where $\gamma_1=1\times 10^{-5}$ and $n\geq 1$. For all three cases, $\gamma_1=1\times 10^{-5}$, $\tau_n = 0.1$, $t_1 = 8.0\;s$, $\Delta t_n = t_n - (t_{n-1}+\tau_n)$, $n\geq 1$}
\end{figure}

Now by varying the magnitude of the decrease $\Delta\gamma$ in the coupling constant $\gamma_n$, we can control the steady - state value which $\left\langle\hat{S}_x (t)\right\rangle$ approaches over time, as well as the interval of time that elapses before the amplitude of the oscillations have decreased sufficiently such that $\left\langle\hat{S}_x (t)\right\rangle$ can be definitively said to be approaching its steady state value. This is shown in figure \ref{fig:spinxexpvalvaryinggamma}. As $\Delta\gamma$ decreases, the steady - state value that $\left\langle\hat{S}_x (t)\right\rangle$ approaches decreases, but so too does the interval of time over which the amplitude of oscillations of $\left\langle\hat{S}_x (t)\right\rangle$ have decreased sufficiently so that it begins to approach its steady - state value.

At this point, we have shown that evolving the trapped ultracold atom gas stroboscopically, with the coupling constant $\gamma$ being decreased each time the coupling is re-established, would result in the suppression of the oscillatory behavior of $\left\langle\hat{S}_x (t)\right\rangle$ and allow it to approach a steady - state value over time. However, there is a question of how one can achieve this stroboscopic time evolution of the trapped atom gas. To answer this question, let us recall the explicit form of the coupling constant $\gamma$, given by Eq. \ref{gammacoeff}. It can be seen from this expression that $\gamma$ is directly proportional to the Rabi frequency $\Omega$ of the laser that couples the ground state in the double wells to the excited state in the harmonic potential in which the double wells are embedded. As such, it is possible for us to reduce the magnitude of the coupling constant $\gamma$ by reducing the laser's Rabi frequency until the desired value is attained. 

\begin{figure}[htb]
\includegraphics[width=0.5\columnwidth, height=0.2\textheight]{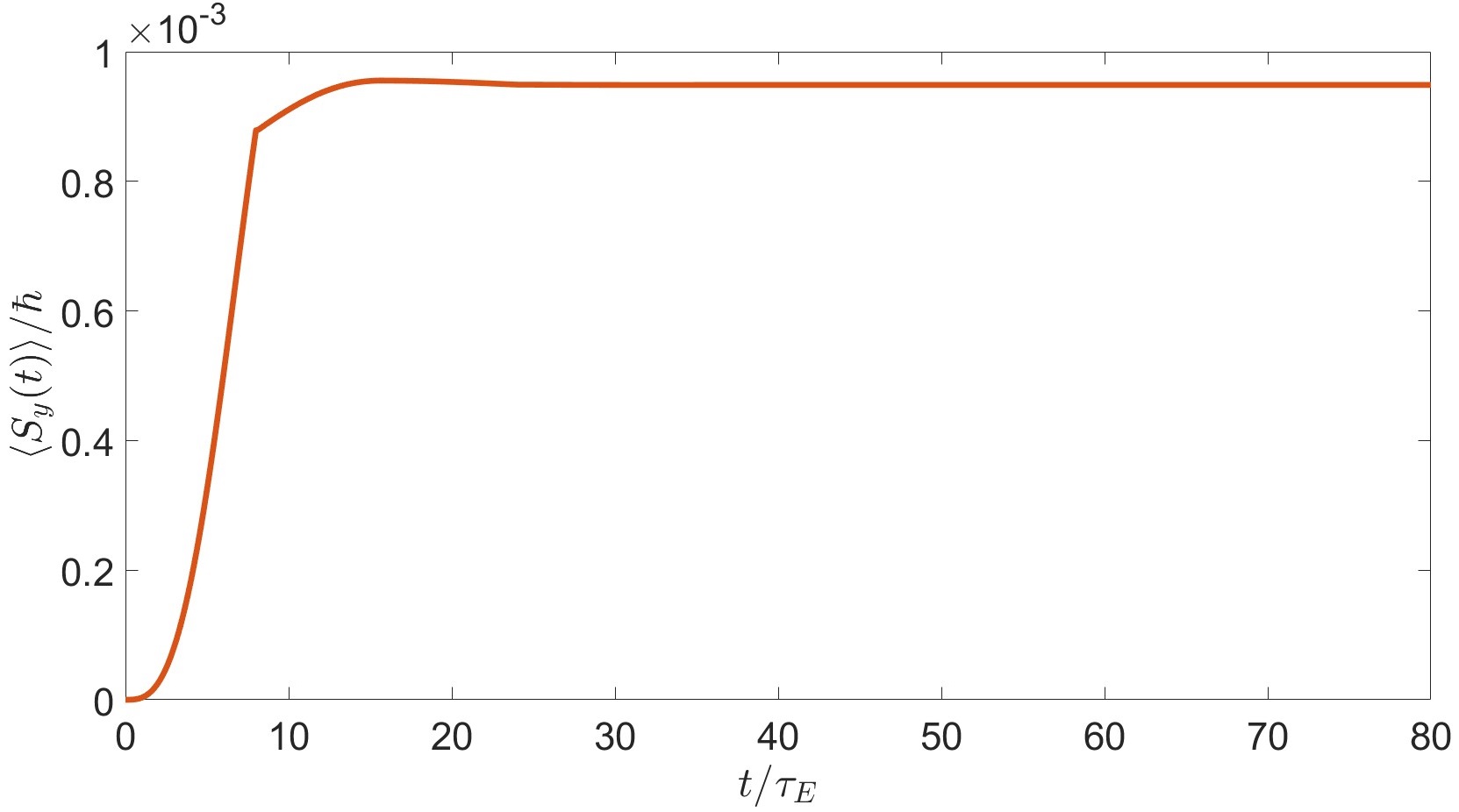}
\includegraphics[width=0.5\columnwidth, height=0.2\textheight]{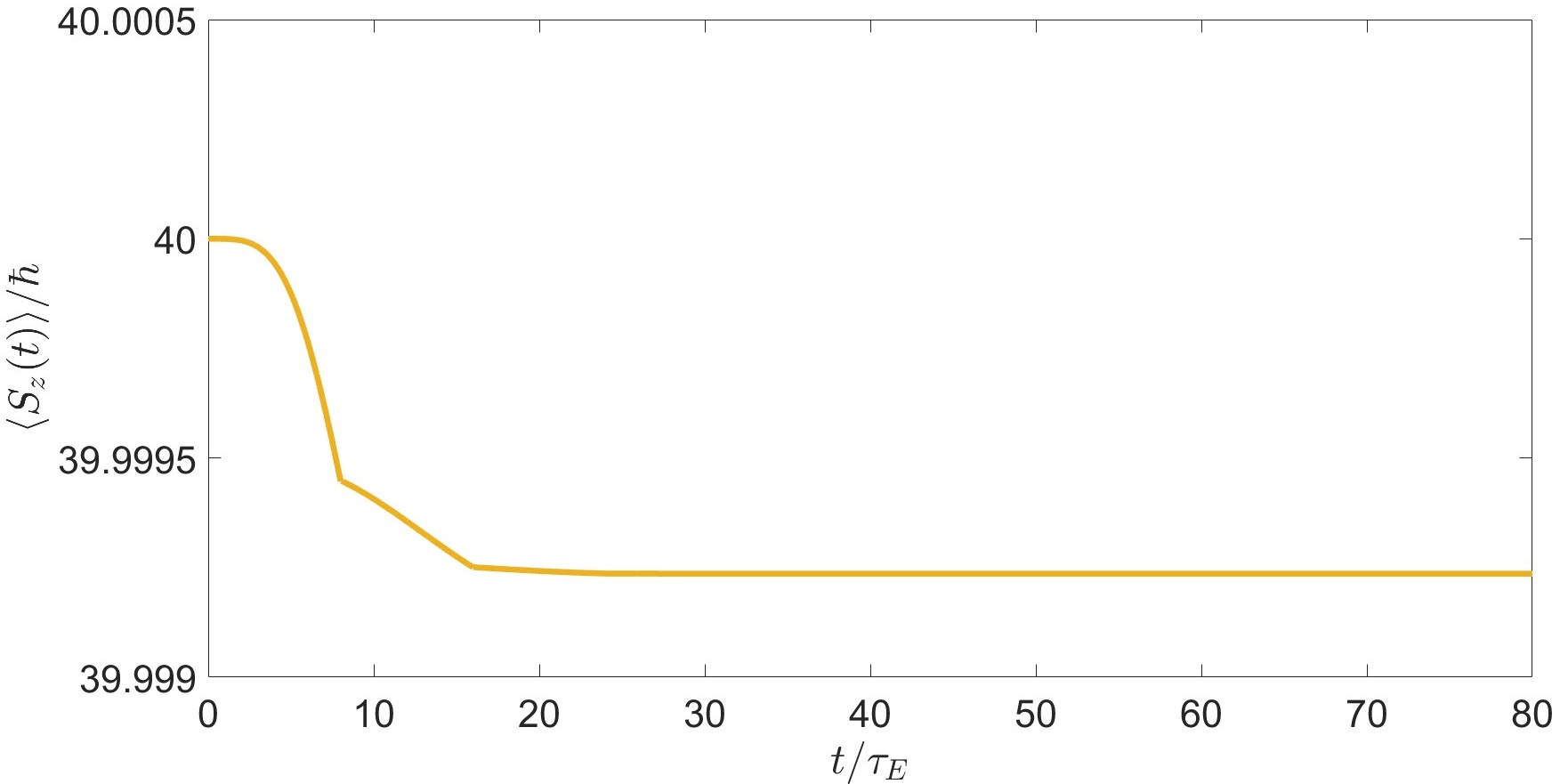}
\caption{\label{fig:spinexpvalvardiag} Time - evolved expectation value of the $SU(2)$ generators $\hat{S}_y(t)$ (top) and $\hat{S}_z$ (bottom), calculated using the time - evolved states $\left|\psi(t)\right\rangle$, with $u=v=2$ for these states and with $\gamma$ varying over time as $\gamma_{n}=0.1\gamma_{n-1}$, $\gamma_1=1\times 10^{-5}$, $\tau_n = 0.1$, $t_1 = 8.0$, $\Delta t_n = t_n - (t_{n-1}+\tau_1)$, $n\geq 1$}
\end{figure}

We note that while the stroboscopic method of evolving the system suppresses the oscillations of the expectation value of $\hat{S}_x (t)$, it also has a similar effect for both $\hat{S}_y (t)$ and $\hat{S}_z (t)$, as shown in figure \ref{fig:spinexpvalvardiag}. In particular, comparing this with figure \ref{fig:spinexpvaldiag}, we find that the oscillations in both $\hat{S}_x (t)$ and $\hat{S}_y (t)$ are damped, allowing the expectation values of both observables to evolve towards a steady - state value. On the other hand, instead of increasing over time, $\hat{S}_z (t)$ is now evolving in a manner similar to both $\hat{S}_x (t)$ and $\hat{S}_y (t)$, approaching a steady - state value as it continues to evolve over time. Therefore, we can say that coupling a trapped ultracold atom gas with a background BEC that acts as a reservoir of excitations and turning this coupling on and off with decreasing strength as the system evolves over time will result in the trapped ultracold atom gas becoming a system whose spin approaches a steady - state value. 

\section{Conclusion}

We have shown, in this paper, that a gas of ultracold bosonic atoms trapped in a double - well potential embedded in a wide harmonic potential which in turn is coupled to a background BEC that acts as a reservoir of excitations will evolve towards a total spin steady state, as evidenced by the expectation value of these two observables for this trapped ultracold bosonic atom gas approaching a constant value over time. However, this steady state can only be achieved by stroboscopic coupling between the trapped ultracold atom gas and the background BEC, wherein the coupling between the system and the environment is turned on and off over fixed time intervals, with the coupling strength decreasing each time it is turned on. 

The resulting steady state from this dissipative quantum state preparation mechanism is of significance not just in many - body algorithms for quantum computing that can be used to simulate quantum chemistry processes \cite{mcclean,bauer}, but also for simulation of quantum many - body systems \cite{georgescu} such as the Hubbard model and spin models, for which definite values of the total spin of the system are necessary. At the same time, considering that the steady states resulting from this dissipative quantum state mechanism have definite spin, one can use this for many - body spintronic applications, such as those described in Refs.\cite{phuc,bello}. However, for this to be experimentally feasible, the features shown in this paper must also be seen if the number of bosonic atoms in the trapped ultracold atom gas are increased by at least one order of magnitude, which corresponds to the standard number of atoms that are present in a BEC. This, together with the determination of the appropriate parameters for the system such as the interatomic scattering length and species of atoms to be used, will be the subject of future work.

Finally, we would like to note that the expectation value of the x - component of the spin, $\hat{S}_x$, shows significant variations over time as the system evolves according to Eq. \ref{masteqevo4}, while the expectation values of the y - and z - components of the spin, given by the operators $\hat{S}_y$ and $\hat{S}_z$ respectively, do not show any significant variation over time. This dynamical behavior is of interest, since it shows that only one component of the spin varies significantly over time while the other two do not. Currently, the author cannot explain why this is the case, and formulating the correct theory that would explain this dynamical behavior will be another subject of future work emerging from this research. 

\section*{Acknowledgments}
The author would like to acknowledge support from the Department of Mathematics and Physics of the College of Science, and the Research Center for Natural and Applied Sciences, of the University of Santo Tomas. The author would also like to acknowledge invaluable comments and suggestions from V. P. Villegas and T. B. O. Tejada. 

\appendix

\section{Evaluation of the Commutators and Tracing Out the Background BEC Observables of the Time - Evolved Interaction Hamiltonian in the Master Equation}
Let us evaluate the explicit form of the commutators appearing in the master equation given by Eq. \ref{masteq}, afterwhich we trace out the background BEC observables, specifically the BEC creation and annihilation operators. In evaluating these commutators and tracing out these observables, we make use of the approximations $\text{Tr}(\hat{b}_k \hat{b}_k' \hat{R})=\text{Tr}(\hat{b}_k^\dagger \hat{b}_k'^\dagger \hat{R})=0$, $\text{Tr}(\hat{b}_k^\dagger \hat{b}_k' \hat{R})=0$ and $\text{Tr}(\hat{b}_k \hat{b}_k'^\dagger \hat{R})\approx \delta_{k,k'}$. We start with the double commutators involving both of the first terms of the time - evolved interaction Hamiltonian:
\begin{eqnarray}
&&\text{Tr}_{\hat{R}}\sum_{k,k'}\sqrt{kk'}\left[(\hat{a}_{g,+}^{\dagger}\hat{a}_{g,+}+\hat{a}_{g,-}^{\dagger}\hat{a}_{g,-})(e^{-\frac{it}{\hbar}E_k}\hat{b}_k +e^{\frac{it}{\hbar}E_k}\hat{b}_k^\dagger),[(\hat{a}_{g,+}^{\dagger}\hat{a}_{g,+}+\hat{a}_{g,-}^{\dagger}\hat{a}_{g,-})\right.\nonumber\\
&&\left.\times(e^{-\frac{i(t-t')}{\hbar}E_k'}\hat{b}_k' +e^{\frac{i(t-t')}{\hbar}E_k'}\hat{b}_k'^\dagger),\hat{\rho}\otimes\hat{R}]\right]\nonumber\\
&&=\sum_{k}k\:e^{-\frac{it'}{\hbar}E_k}\left((\hat{a}_{g,+}^{\dagger}\hat{a}_{g,+}+\hat{a}_{g,-}^{\dagger}\hat{a}_{g,-})(\hat{a}_{g,+}^{\dagger}\hat{a}_{g,+}+\hat{a}_{g,-}^{\dagger}\hat{a}_{g,-})\hat{\rho}\right.\nonumber\\
&&\left.-(\hat{a}_{g,+}^{\dagger}\hat{a}_{g,+}+\hat{a}_{g,-}^{\dagger}\hat{a}_{g,-})\hat{\rho}(\hat{a}_{g,+}^{\dagger}\hat{a}_{g,+}+\hat{a}_{g,-}^{\dagger}\hat{a}_{g,-})\right)\nonumber\\
&&-\sum_{k}k\:e^{\frac{it'}{\hbar}E_k}\left((\hat{a}_{g,+}^{\dagger}\hat{a}_{g,+}+\hat{a}_{g,-}^{\dagger}\hat{a}_{g,-})\hat{\rho}(\hat{a}_{g,+}^{\dagger}\hat{a}_{g,+}+\hat{a}_{g,-}^{\dagger}\hat{a}_{g,-})\right.\nonumber\\
&&\left.-\hat{\rho}(\hat{a}_{g,+}^{\dagger}\hat{a}_{g,+}+\hat{a}_{g,-}^{\dagger}\hat{a}_{g,-})(\hat{a}_{g,+}^{\dagger}\hat{a}_{g,+}+\hat{a}_{g,-}^{\dagger}\hat{a}_{g,-})\right)\nonumber\\
\label{masteqterm1prov}
\end{eqnarray}
Next, we evaluate the double commutators in the master equation involving the first and the second terms of the time - evolved interaction Hamiltonian, obtaining the following expression:
\begin{eqnarray}
&&2\sqrt{\frac{m_S \omega_{e,x}}{\hbar}}\left(\frac{\omega_{e,x}}{\omega_{g,x}}\right)^{1/4}x_0 e^{-\frac{m_s\omega_e}{2\hbar} x_0^2}\text{Tr}_{\hat{R}}\sum_{k,k'}\sqrt{kk'}\left[(\hat{a}_{g,+}^{\dagger}\hat{a}_{g,+}+\hat{a}_{g,-}^{\dagger}\hat{a}_{g,-})(e^{-\frac{it}{\hbar}E_k}\hat{b}_k +e^{\frac{it}{\hbar}E_k}\hat{b}_k^\dagger),\right.\nonumber\\
&&[e^{-\frac{i(t-t')}{\hbar}(E_{k'} +(\epsilon_e -\epsilon_g))}(\hat{a}^{\dagger}_{g,+}-\hat{a}^{\dagger}_{g,-})\hat{a}_{e,0}\hat{b}_{k'}+e^{\frac{i(t-t')}{\hbar}(E_{k'} -(\epsilon_e -\epsilon_g))}(\hat{a}^{\dagger}_{g,+}-\hat{a}^{\dagger}_{g,-})\hat{a}_{e,0}\hat{b}^{\dagger}_{k'}\nonumber\\
&&\left.+e^{-\frac{i(t-t')}{\hbar}(E_{k'} -(\epsilon_e -\epsilon_g))}(\hat{a}_{g,+}-\hat{a}_{g,-})\hat{a}^{\dagger}_{e,0}\hat{b}_{k'}+e^{\frac{i(t-t')}{\hbar}(E_{k'} +(\epsilon_e -\epsilon_g))}(\hat{a}_{g,+}-\hat{a}_{g,-})\hat{a}^{\dagger}_{e,0}\hat{b}^{\dagger}_{k'},\hat{\rho}\otimes\hat{R}]\right]\nonumber\\
&&=2\sqrt{\frac{m_S \omega_{e,x}}{\hbar}}\left(\frac{\omega_{e,x}}{\omega_{g,x}}\right)^{1/4}x_0 e^{-\frac{m_s\omega_e}{2\hbar} x_0^2}\sum_{k}k\left(e^{-\frac{it}{\hbar}(\epsilon_e -\epsilon_g)}e^{-\frac{it'}{\hbar}(E_{k}-(\epsilon_e -\epsilon_g))}\right.\nonumber\\
&&\times((\hat{a}_{g,+}^{\dagger}\hat{a}_{g,+}+\hat{a}_{g,-}^{\dagger}\hat{a}_{g,-})(\hat{a}^{\dagger}_{g,+}-\hat{a}^{\dagger}_{g,-})\hat{a}_{e,0}\hat{\rho}-(\hat{a}^{\dagger}_{g,+}-\hat{a}^{\dagger}_{g,-})\hat{a}_{e,0}\hat{\rho}(\hat{a}_{g,+}^{\dagger}\hat{a}_{g,+}+\hat{a}_{g,-}^{\dagger}\hat{a}_{g,-}))\nonumber\\
&&+e^{\frac{it}{\hbar}(\epsilon_e -\epsilon_g)}e^{-\frac{it'}{\hbar}(E_{k}+(\epsilon_e -\epsilon_g))}((\hat{a}_{g,+}^{\dagger}\hat{a}_{g,+}+\hat{a}_{g,-}^{\dagger}\hat{a}_{g,-})(\hat{a}_{g,+}-\hat{a}_{g,-})\hat{a}^{\dagger}_{e,0}\hat{\rho}\nonumber\\
&&\left.-(\hat{a}_{g,+}-\hat{a}_{g,-})\hat{a}^{\dagger}_{e,0}\hat{\rho}(\hat{a}_{g,+}^{\dagger}\hat{a}_{g,+}+\hat{a}_{g,-}^{\dagger}\hat{a}_{g,-}))\right)\nonumber\\
&&-2\sqrt{\frac{m_S \omega_{e,x}}{\hbar}}\left(\frac{\omega_{e,x}}{\omega_{g,x}}\right)^{1/4}x_0 e^{-\frac{m_s\omega_e}{2\hbar} x_0^2}\sum_{k}k\left(e^{-\frac{it}{\hbar}(\epsilon_e -\epsilon_g)}e^{\frac{it'}{\hbar}(E_{k}+(\epsilon_e -\epsilon_g))}\right.\nonumber\\
&&\times((\hat{a}_{g,+}^{\dagger}\hat{a}_{g,+}+\hat{a}_{g,-}^{\dagger}\hat{a}_{g,-})\hat{\rho}(\hat{a}^{\dagger}_{g,+}-\hat{a}^{\dagger}_{g,-})\hat{a}_{e,0}-\hat{\rho}(\hat{a}^{\dagger}_{g,+}-\hat{a}^{\dagger}_{g,-})\hat{a}_{e,0}(\hat{a}_{g,+}^{\dagger}\hat{a}_{g,+}+\hat{a}_{g,-}^{\dagger}\hat{a}_{g,-}))\nonumber\\
&&+e^{\frac{it}{\hbar}(\epsilon_e -\epsilon_g)}e^{\frac{it'}{\hbar}(E_{k}-(\epsilon_e -\epsilon_g))}((\hat{a}_{g,+}^{\dagger}\hat{a}_{g,+}+\hat{a}_{g,-}^{\dagger}\hat{a}_{g,-})\hat{\rho}(\hat{a}_{g,+}-\hat{a}_{g,-})\hat{a}^{\dagger}_{e,0}\nonumber\\
&&\left.-\hat{\rho}(\hat{a}_{g,+}-\hat{a}_{g,-})\hat{a}^{\dagger}_{e,0}(\hat{a}_{g,+}^{\dagger}\hat{a}_{g,+}+\hat{a}_{g,-}^{\dagger}\hat{a}_{g,-}))\right)\nonumber\\
\label{masteqterm2prov}
\end{eqnarray}
Evaluating the double commutators in the master equation involving the second and the first terms of the time - evolved interaction Hamiltonian, we obtain the following expression:
\begin{eqnarray}
&&2\sqrt{\frac{m_S \omega_{e,x}}{\hbar}}\left(\frac{\omega_{e,x}}{\omega_{g,x}}\right)^{1/4}e^{-\frac{m_s\omega_e}{2\hbar} x_0^2}x_0\;\text{Tr}_{\hat{R}}\sum_{k,k'}\sqrt{kk'}\left[e^{-\frac{it}{\hbar}(E_{k} +(\epsilon_e -\epsilon_g))}(\hat{a}^{\dagger}_{g,+}-\hat{a}^{\dagger}_{g,-})\hat{a}_{e,0}\hat{b}_{k}\right.\nonumber\\
&&+e^{\frac{it}{\hbar}(E_{k} -(\epsilon_e -\epsilon_g))}(\hat{a}^{\dagger}_{g,+}-\hat{a}^{\dagger}_{g,-})\hat{a}_{e,0}\hat{b}^{\dagger}_{k}+e^{-\frac{it}{\hbar}(E_{k} -(\epsilon_e -\epsilon_g))}(\hat{a}_{g,+}-\hat{a}_{g,-})\hat{a}^{\dagger}_{e,0}\hat{b}_{k}\nonumber\\
&&\left.+e^{\frac{it}{\hbar}(E_{k} +(\epsilon_e -\epsilon_g))}(\hat{a}_{g,+}-\hat{a}_{g,-})\hat{a}^{\dagger}_{e,0}\hat{b}^{\dagger}_{k'},[(\hat{a}_{g,+}^{\dagger}\hat{a}_{g,+}+\hat{a}_{g,-}^{\dagger}\hat{a}_{g,-})(e^{-\frac{i(t-t')}{\hbar}E_{k'}}\hat{b}_{k'}+e^{\frac{i(t-t')}{\hbar}E_{k'}}\hat{b}_{k'}^\dagger),\hat{\rho}\otimes\hat{R}]\right]\nonumber\\
&&=2\sqrt{\frac{m_S \omega_{e,x}}{\hbar}}\left(\frac{\omega_{e,x}}{\omega_{g,x}}\right)^{1/4}e^{-\frac{m_s\omega_e}{2\hbar} x_0^2}x_0\nonumber\\
&&\times\sum_{k}k\left(e^{-\frac{it'}{\hbar}E_{k}}e^{-\frac{it}{\hbar}(\epsilon_e -\epsilon_g)}((\hat{a}^{\dagger}_{g,+}-\hat{a}^{\dagger}_{g,-})\hat{a}_{e,0}(\hat{a}_{g,+}^{\dagger}\hat{a}_{g,+}+\hat{a}_{g,-}^{\dagger}\hat{a}_{g,-})\hat{\rho}\right.\nonumber\\
&&-(\hat{a}_{g,+}^{\dagger}\hat{a}_{g,+}+\hat{a}_{g,-}^{\dagger}\hat{a}_{g,-})\hat{\rho}(\hat{a}^{\dagger}_{g,+}-\hat{a}^{\dagger}_{g,-})\hat{a}_{e,0})\nonumber\\
&&+e^{-\frac{it'}{\hbar}E_{k}}e^{\frac{it}{\hbar}(\epsilon_e -\epsilon_g)}((\hat{a}_{g,+}-\hat{a}_{g,-})\hat{a}^{\dagger}_{e,0}(\hat{a}_{g,+}^{\dagger}\hat{a}_{g,+}+\hat{a}_{g,-}^{\dagger}\hat{a}_{g,-})\hat{\rho}\nonumber\\
&&\left.-(\hat{a}_{g,+}^{\dagger}\hat{a}_{g,+}+\hat{a}_{g,-}^{\dagger}\hat{a}_{g,-})\hat{\rho}(\hat{a}_{g,+}-\hat{a}_{g,-})\hat{a}^{\dagger}_{e,0})\right)\nonumber\\
&&-2\sqrt{\frac{m_S \omega_{e,x}}{\hbar}}\left(\frac{\omega_{e,x}}{\omega_{g,x}}\right)^{1/4}e^{-\frac{m_s\omega_e}{2\hbar} x_0^2}x_0\nonumber\\
&&\times\sum_{k}k\left(e^{\frac{it'}{\hbar}E_{k}}e^{-\frac{it}{\hbar}(\epsilon_e -\epsilon_g)}((\hat{a}^{\dagger}_{g,+}-\hat{a}^{\dagger}_{g,-})\hat{a}_{e,0}\hat{\rho}(\hat{a}_{g,+}^{\dagger}\hat{a}_{g,+}+\hat{a}_{g,-}^{\dagger}\hat{a}_{g,-})\right.\nonumber\\
&&-\hat{\rho}(\hat{a}_{g,+}^{\dagger}\hat{a}_{g,+}+\hat{a}_{g,-}^{\dagger}\hat{a}_{g,-})(\hat{a}^{\dagger}_{g,+}-\hat{a}^{\dagger}_{g,-})\hat{a}_{e,0})\nonumber\\
&&+e^{\frac{it'}{\hbar}E_{k}}e^{\frac{it}{\hbar}(\epsilon_e -\epsilon_g)}((\hat{a}_{g,+}-\hat{a}_{g,-})\hat{a}^{\dagger}_{e,0}\hat{\rho}(\hat{a}_{g,+}^{\dagger}\hat{a}_{g,+}+\hat{a}_{g,-}^{\dagger}\hat{a}_{g,-})\nonumber\\
&&\left.-\hat{\rho}(\hat{a}_{g,+}^{\dagger}\hat{a}_{g,+}+\hat{a}_{g,-}^{\dagger}\hat{a}_{g,-})(\hat{a}_{g,+}-\hat{a}_{g,-})\hat{a}^{\dagger}_{e,0})\right)\nonumber\\
\label{masteqterm3prov}
\end{eqnarray}
Finally, evaluating the double commutators in the master equation involving both of the last terms in the time - evolved interaction Hamiltonian, we obtain the following expression:
\begin{eqnarray}
&&4\frac{m_S \omega_{e,x}}{\hbar}\sqrt{\frac{\omega_{e,x}}{\omega_{g,x}}}e^{-\frac{m_s\omega_e}{\hbar} x_0^2} x^{2}_0\sum_{k,k'}\sqrt{kk'}\text{Tr}_{\hat{R}}\left[e^{-\frac{it}{\hbar}(E_k +(\epsilon_e -\epsilon_g))}(\hat{a}^{\dagger}_{g,+}-\hat{a}^{\dagger}_{g,-})\hat{a}_{e,0}\hat{b}_k\right.\nonumber\\
&&+e^{\frac{it}{\hbar}(E_k -(\epsilon_e -\epsilon_g))}(\hat{a}^{\dagger}_{g,+}-\hat{a}^{\dagger}_{g,-})\hat{a}_{e,0}\hat{b}^{\dagger}_k+e^{-\frac{it}{\hbar}(E_k -(\epsilon_e -\epsilon_g))}(\hat{a}_{g,+}-\hat{a}_{g,-})\hat{a}^{\dagger}_{e,0}\hat{b}_k\nonumber\\
&&+e^{\frac{it}{\hbar}(E_k +(\epsilon_e -\epsilon_g))}(\hat{a}_{g,+}-\hat{a}_{g,-})\hat{a}^{\dagger}_{e,0}\hat{b}^{\dagger}_k,\nonumber\\
&&[e^{-\frac{i(t-t')}{\hbar}(E_{k'} +(\epsilon_e -\epsilon_g))}(\hat{a}^{\dagger}_{g,+}-\hat{a}^{\dagger}_{g,-})\hat{a}_{e,0}\hat{b}_{k'}\nonumber\\
&&+e^{\frac{i(t-t')}{\hbar}(E_{k'}-(\epsilon_e -\epsilon_g))}(\hat{a}^{\dagger}_{g,+}-\hat{a}^{\dagger}_{g,-})\hat{a}_{e,0}\hat{b}^{\dagger}_{k'}+e^{-\frac{i(t-t')}{\hbar}(E_{k'} -(\epsilon_e -\epsilon_g))}(\hat{a}_{g,+}-\hat{a}_{g,-})\hat{a}^{\dagger}_{e,0}\hat{b}_{k'}\nonumber\\
&&\left.+e^{\frac{i(t-t')}{\hbar}(E_{k'}+(\epsilon_e -\epsilon_g))}(\hat{a}_{g,+}-\hat{a}_{g,-})\hat{a}^{\dagger}_{e,0}\hat{b}^{\dagger}_{k'},\hat{\rho}\otimes\hat{R}]\right]\nonumber\\
&&=4\frac{m_S \omega_{e,x}}{\hbar}\sqrt{\frac{\omega_{e,x}}{\omega_{g,x}}}e^{-\frac{m_s\omega_e}{\hbar} x_0^2} x^{2}_0\nonumber\\
&&\times\sum_{k}k\left(e^{-\frac{2it}{\hbar}(\epsilon_e -\epsilon_g)}e^{-\frac{it'}{\hbar}(E_{k} -(\epsilon_e -\epsilon_g))}((\hat{a}^{\dagger}_{g,+}-\hat{a}^{\dagger}_{g,-})\hat{a}_{e,0}(\hat{a}^{\dagger}_{g,+}-\hat{a}^{\dagger}_{g,-})\hat{a}_{e,0}\hat{\rho}\right.\nonumber\\
&&-(\hat{a}^{\dagger}_{g,+}-\hat{a}^{\dagger}_{g,-})\hat{a}_{e,0}\hat{\rho}(\hat{a}^{\dagger}_{g,+}-\hat{a}^{\dagger}_{g,-})\hat{a}_{e,0})\nonumber\\
&&+e^{-\frac{it'}{\hbar}(E_{k}+(\epsilon_e -\epsilon_g))}((\hat{a}^{\dagger}_{g,+}-\hat{a}^{\dagger}_{g,-})\hat{a}_{e,0}(\hat{a}_{g,+}-\hat{a}_{g,-})\hat{a}^{\dagger}_{e,0}\hat{\rho}-(\hat{a}_{g,+}-\hat{a}_{g,-})\hat{a}^{\dagger}_{e,0}\hat{\rho}(\hat{a}^{\dagger}_{g,+}-\hat{a}^{\dagger}_{g,-})\hat{a}_{e,0})\nonumber\\
&&+e^{-\frac{it'}{\hbar}(E_{k}-(\epsilon_e -\epsilon_g))}((\hat{a}_{g,+}-\hat{a}_{g,-})\hat{a}^{\dagger}_{e,0}(\hat{a}^{\dagger}_{g,+}-\hat{a}^{\dagger}_{g,-})\hat{a}_{e,0}\hat{\rho}-(\hat{a}^{\dagger}_{g,+}-\hat{a}^{\dagger}_{g,-})\hat{a}_{e,0}\hat{\rho}(\hat{a}_{g,+}-\hat{a}_{g,-})\hat{a}^{\dagger}_{e,0})\nonumber\\
&&+e^{\frac{2it}{\hbar}(\epsilon_e -\epsilon_g)}e^{-\frac{it'}{\hbar}(E_{k}+(\epsilon_e -\epsilon_g))}((\hat{a}_{g,+}-\hat{a}_{g,-})\hat{a}^{\dagger}_{e,0}(\hat{a}_{g,+}-\hat{a}_{g,-})\hat{a}^{\dagger}_{e,0}\hat{\rho}\nonumber\\
&&\left.-(\hat{a}_{g,+}-\hat{a}_{g,-})\hat{a}^{\dagger}_{e,0}\hat{\rho}(\hat{a}_{g,+}-\hat{a}_{g,-})\hat{a}^{\dagger}_{e,0})\right)\nonumber\\
&&-4\frac{m_S \omega_{e,x}}{\hbar}\sqrt{\frac{\omega_{e,x}}{\omega_{g,x}}}e^{-\frac{m_s\omega_e}{\hbar} x_0^2} x^{2}_0\nonumber\\
&&\times\sum_{k}k\left(e^{-\frac{2it}{\hbar}(\epsilon_e -\epsilon_g)}e^{\frac{it'}{\hbar}(E_{k}+(\epsilon_e -\epsilon_g))}((\hat{a}^{\dagger}_{g,+}-\hat{a}^{\dagger}_{g,-})\hat{a}_{e,0}\hat{\rho}(\hat{a}^{\dagger}_{g,+}-\hat{a}^{\dagger}_{g,-})\hat{a}_{e,0}\right.\nonumber\\
&&-(\hat{a}^{\dagger}_{g,+}-\hat{\rho}\hat{a}^{\dagger}_{g,-})\hat{a}_{e,0}(\hat{a}^{\dagger}_{g,+}-\hat{a}^{\dagger}_{g,-})\hat{a}_{e,0})\nonumber\\
&&+e^{\frac{it'}{\hbar}(E_{k}-(\epsilon_e -\epsilon_g))}((\hat{a}^{\dagger}_{g,+}-\hat{a}^{\dagger}_{g,-})\hat{a}_{e,0}\hat{\rho}(\hat{a}_{g,+}-\hat{a}_{g,-})\hat{a}^{\dagger}_{e,0}-\hat{\rho}(\hat{a}_{g,+}-\hat{a}_{g,-})\hat{a}^{\dagger}_{e,0}(\hat{a}^{\dagger}_{g,+}-\hat{a}^{\dagger}_{g,-})\hat{a}_{e,0})\nonumber\\
&&+e^{\frac{it'}{\hbar}(E_{k}+(\epsilon_e -\epsilon_g))}((\hat{a}_{g,+}-\hat{a}_{g,-})\hat{a}^{\dagger}_{e,0}\hat{\rho}(\hat{a}^{\dagger}_{g,+}-\hat{a}^{\dagger}_{g,-})\hat{a}_{e,0}-\hat{\rho}(\hat{a}^{\dagger}_{g,+}-\hat{a}^{\dagger}_{g,-})\hat{a}_{e,0}(\hat{a}_{g,+}-\hat{a}_{g,-})\hat{a}^{\dagger}_{e,0})\nonumber\\
&&+e^{\frac{2it}{\hbar}(\epsilon_e -\epsilon_g)}e^{\frac{it'}{\hbar}(E_{k}-(\epsilon_e -\epsilon_g))}((\hat{a}_{g,+}-\hat{a}_{g,-})\hat{a}^{\dagger}_{e,0}\hat{\rho}(\hat{a}_{g,+}-\hat{a}_{g,-})\hat{a}^{\dagger}_{e,0}\nonumber\\
&&\left.-\hat{\rho}(\hat{a}_{g,+}-\hat{a}_{g,-})\hat{a}^{\dagger}_{e,0}(\hat{a}_{g,+}-\hat{a}_{g,-})\hat{a}^{\dagger}_{e,0})\right)\nonumber\\
\label{masteqterm4prov}
\end{eqnarray}
\bibliographystyle{elsarticle-num-names}

\bibliography{DBEC}

\end{document}